\documentclass[%
 reprint,
superscriptaddress,
 amsmath,amssymb,
 aps,
pra,
floatfix,
]{revtex4-2}
\usepackage[utf8]{inputenc}
\usepackage{booktabs}
\usepackage{graphicx}% Include figure files
\usepackage{dcolumn}% Align table columns on decimal point
\usepackage{bm}% bold math
\usepackage{hyperref}
\usepackage{amsfonts}
\usepackage{algorithmic}
\usepackage{textcomp}
\usepackage{xcolor}
\usepackage{multirow}
\usepackage{braket}
\usepackage{quantikz}
\usepackage{tikz}
\usepackage{graphicx}
\usepackage{totpages}
\usepackage{cleveref}
\usepackage{siunitx}
\renewcommand{\selectlanguage}[1]{} % RevTex 4-2: Package babel Error: You haven't defined the language en yet

\Crefname{equation}{Eq.}{Eqs.}
\crefname{equation}{Eq.}{Eqs.}
\Crefname{definition}{Def.}{Defs.}
\crefname{definition}{Def.}{Defs.}
\Crefname{figure}{Fig.}{Figs.}
\crefname{figure}{Fig.}{Figs.}
\Crefname{section}{Sec.}{Secs.}
\crefname{section}{Sec.}{Secs.}
\Crefname{assumption}{Assum.}{Assums.}
\crefname{assumption}{Assum.}{Assums.}

\AddToHook{cmd/appendix/before}{%
  \crefalias{section}{appendix}%
  \crefalias{subsection}{subappendix}%
  \crefalias{subsubsection}{subsubappendix}%
}
\begin{document}

\preprint{APS/123-QED}

\title{Assessing System Capabilities and Bottlenecks of an Early Fault-Tolerant Bicycle Architecture}

\author{Kun Liu}
\email{kun.liu.kl944@yale.edu}
\affiliation{Department of Computer Science, Yale University, New Haven, Connecticut 06511, USA}
\affiliation{Yale Quantum Institute, Yale University, New Haven, Connecticut 06511, USA}

\author{Ben Foxman}
\affiliation{Department of Computer Science, Yale University, New Haven, Connecticut 06511, USA}
\affiliation{Yale Quantum Institute, Yale University, New Haven, Connecticut 06511, USA}

\author{Gian-Luca R. Anselmetti}
\affiliation{Quantum Lab, Boehringer Ingelheim, 55218 Ingelheim am Rhein, Germany}

\author{Yongshan Ding}
\email{yongshan.ding@yale.edu}
\affiliation{Department of Computer Science, Yale University, New Haven, Connecticut 06511, USA}
\affiliation{Yale Quantum Institute, Yale University, New Haven, Connecticut 06511, USA}

% \author{Ann Author}
%  \altaffiliation[Also at ]{Physics Department, XYZ University.}%Lines break automatically or can be forced with \\
%  \email{Second.Author@institution.edu}
% \affiliation{%
%  Authors' institution and/or address\\
%  This line break forced with \textbackslash\textbackslash
% }%

% \collaboration{MUSO Collaboration}%\noaffiliation

%  \homepage{http://www.Second.institution.edu/~Charlie.Author}
% \affiliation{
%  Second institution and/or address\\
%  This line break forced% with \\
% }%
% \affiliation{
%  Third institution, the second for Charlie Author
% \affiliation{%
%  Authors' institution and/or address\\
%  This line break forced with \textbackslash\textbackslash
% }%

% \collaboration{CLEO Collaboration}%\noaffiliation

\date{\today}% It is always \today, today,

\begin{abstract}

Early modular fault tolerant quantum computers remain constrained by costly inter-module communication and limited magic state factory service.
Understanding such bottlenecks and investigating compiler optimizations most close the gap between algorithm requirements and hardware capabilities is a concrete and practically urgent systems problem.
We study the modular architectures based on Bivariate Bicycle codes and identify the dominant bottleneck: inter-module communication induced by non-Clifford operations.
We build a compilation pipeline to fill the missing parts of prior works and propose compiler optimizations: synthesizing arbitrary-angle rotations at the factory (syn@fac), transvection based Clifford deferral, and Clifford insertion for critical path duration reduction.
We extend the evaluation scope of the prior work to 40+ benchmark categories drawn from PennyLane and MQTBench, including quantum algorithms and Hamiltonian simulations with varying sizes.
Under the present instruction cost, syn@fac reduces estimated circuit failure probability by a factor of 9.0 on average across non-Clifford benchmarks.
The robustness persists across sweeps of instruction cost ratios, LPU count, and factory count.
Besides, transvection reduces Clifford deferral compile time by 77.04\%, while Clifford insertion reduces end-to-end circuit duration by 11.54\% on average on MQTBench, with smaller gains on Hamiltonian simulations.
We hope this work inspires the studies on compiler optimizations for early modular FTQC systems.

\end{abstract}

\maketitle

\section{Introduction}
\label{sec:intro}

\begin{figure}[t]
    \centering
    \includegraphics[trim={0 0.5cm 0 0}, width=\linewidth]{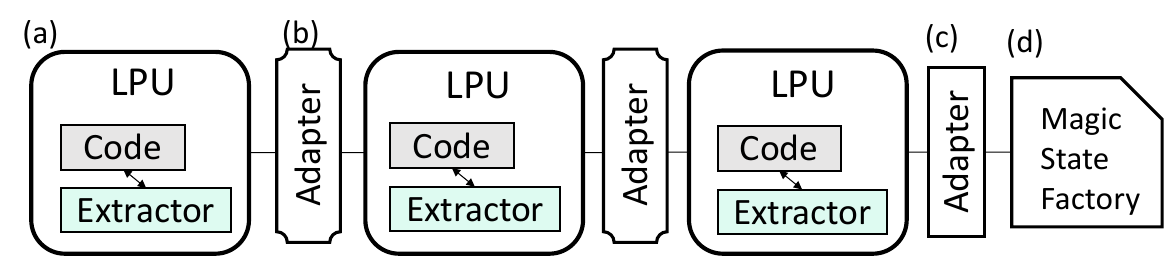}
    \caption{
        The reference modular bicycle architecture, based on Bivariate Bicycle codes proposed in~\cite{yoder_tour_2025}.
        (a) A logical processing unit (LPU) consists of a QEC code plus extractor for arbitrary in-module Pauli measurements.
        (b) Code--code adapter for inter-module Pauli measurements.
        (c) Code--factory adapter for magic state teleportation.
        (d) Magic state factory.
    }
    % \Description{}
    \label{fig:modular_arch}
\end{figure}

\begin{figure*}[htbp]
    \centering
    \includegraphics[trim={0 0.5cm 0 0}, width=0.94\linewidth]{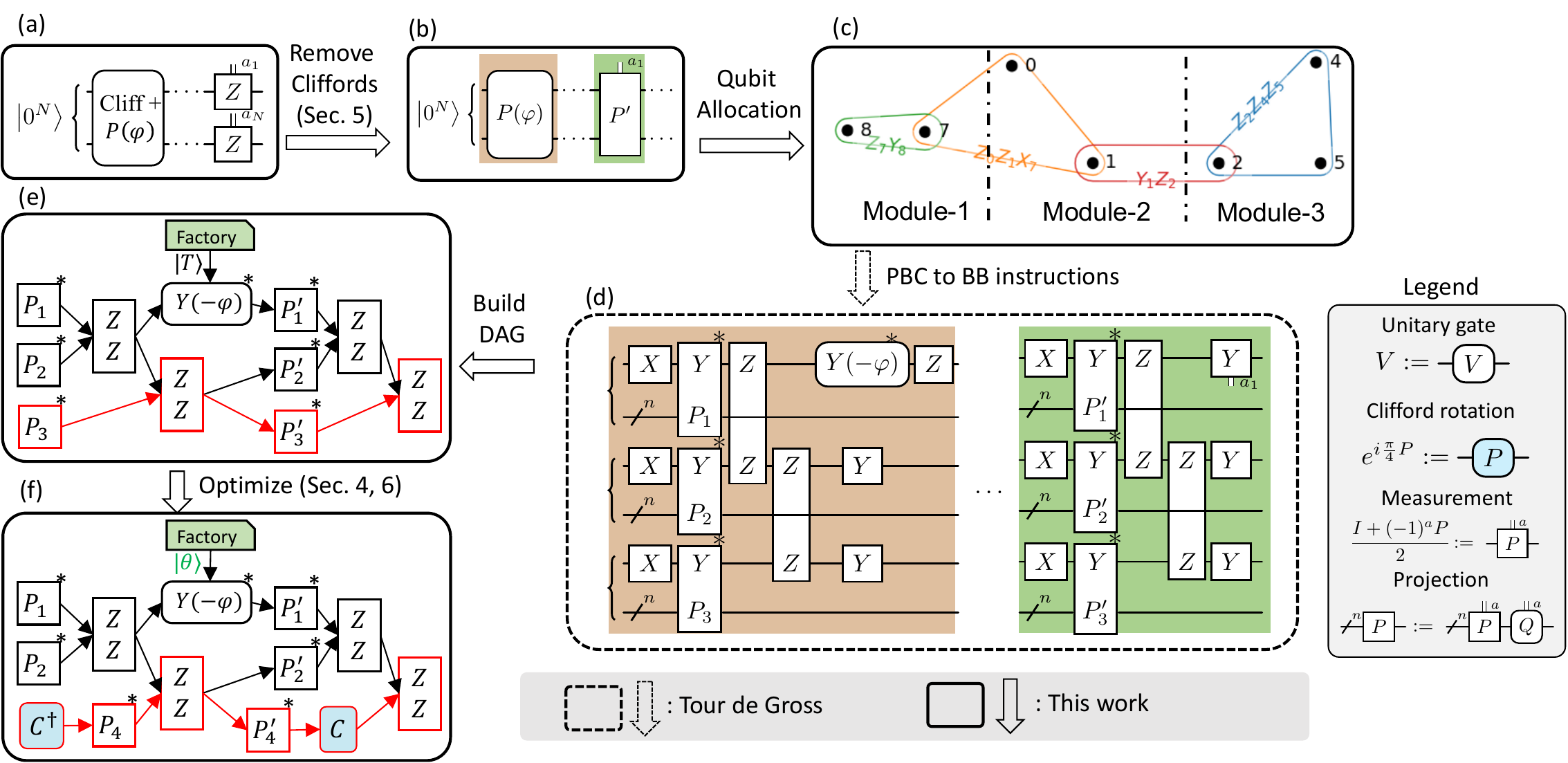}
    \caption{
        Compilation pipeline for the studied systems.
        We adopt the BB instruction set and the PBC-to-BB transformation from Tour de Gross~\cite{yoder_tour_2025}, together with the qubit allocator reproducing same-time work by Sethi \emph{et al.}~\cite{sethi_optimizing_2026}.
        Our main compiler choices intervene at three points:
        Clifford deferral transforms (a) to (b),
        synthesis placement changes how arbitrary-angle rotations are realized after allocation,
        and Clifford insertion optimizes the critical path after DAG construction.
        ``*'' marks a non-native operation requiring further synthesis.
    }
    % \Description{}
    \label{fig:pipeline}
\end{figure*}

\begin{table*}[ht]
    \centering
    \resizebox{\linewidth}{!}{
        \begin{tabular}{llllll}
            \toprule
            Id & Optimization target & Tour de Gross~\cite{yoder_tour_2025} & This work & Effect & Ref. \\
            \midrule
            Opt-1 & Rotation synthesis & syn@LPU & syn@fac & Reduce failure prob., increase duration & \Cref{sec:syn_at_fac} \\
            Opt-2 & Pauli meas. synthesis & N/A & Clifford insertion & Reduce duration  & \Cref{sec:depth_opt} \\
            Opt-3 & Clifford deferral & N/A & Transvection & Reduce compilation runtime & \Cref{sec:transvection} \\
            \bottomrule
            \end{tabular}
    }
    \caption{
        Comparison of the main compiler optimizations of this work and the baseline~\cite{yoder_tour_2025}.
        LPU: logical processing unit.
        syn@LPU: synthesize arbitrary-angle rotations at the LPU.
        syn@fac: synthesize arbitrary-angle rotations at the factory.
    }
    \label{tab:opt_list}
\end{table*}

Early fault-tolerant quantum computers capable of running useful quantum algorithms are beginning to emerge.
They will still be constrained in logical qubit count and bottlenecked by costly fault-tolerant operations, particularly non-Clifford operations.
Understanding which compiler optimizations most close the gap between algorithm requirements and hardware capabilities in this early regime is therefore a concrete and practically urgent systems problem.

Two architectural trends make this early regime plausible.
First, quantum low-density parity check (qLDPC) codes reduce physical qubit overhead, with bivariate bicycle (BB) codes and hypergraph product codes providing concrete high-rate families that substantially lower overhead relative to surface codes
\cite{bravyi_high-threshold_2024, xu_constant-overhead_2023}.
% \cite{krishna_fault-tolerant_2021, panteleev_degenerate_2021, panteleev_quantum_2022, tillich_quantum_2014, kovalev_quantum_2013, bravyi_high-threshold_2024, xu_constant-overhead_2023}.
Second, modular architectures offer a realistic path to scale beyond monolithic chips by partitioning a quantum computer into a small number of quantum error correction (QEC) code blocks connected by sparse inter-module links~\cite{du_hardware-aware_2025, jeng_modular_2025, sang_toward_2025}.

Such early modular fault-tolerant quantum computing (FTQC) architectures face two critical bottlenecks.
The first is an \emph{inter-module communication bottleneck}: fault-tolerantly coupling qLDPC modules requires nontrivial extractor, bridge, gauging, and adapter constructions, and each inter-module Pauli measurement incurs a non-negligible error~\cite{cross_improved_2024, williamson_low-overhead_2024, swaroop_universal_2024}.
The second is a \emph{magic state factory bottleneck}: universal computation still requires magic states for non-Clifford execution, and factory throughput remains finite and costly~\cite{gidney_how_2025, gidney_magic_2024, campbell_efficient_2016}.
As a result, early modular FTQC architectures are likely to be \emph{communication-limited} and \emph{factory-limited}.

Under current logical error rates of the instructions, realistic Pauli-based computation workloads still accumulate large circuit failure probabilities in this early regime.
The compiler question is therefore not how to optimize an ideal modular machine, but which optimizations most reduce that estimated circuit failure probability while respecting the architecture's constraints.
Because the compiler is the primary interface between workloads and hardware constraints, compiler optimization is the primary tool available to improve system capabilities without simply waiting for hardware to improve.
The central question this paper addresses is:
\begin{center}
\fbox{
\parbox{0.95\columnwidth}{
\textbf{Question.}
If early modular FTQC systems are communication-limited and factory-limited, which compiler optimizations most reduce the estimated circuit failure probability of realistic quantum algorithm workloads?
}
}
\end{center}

To make this question concrete and quantifiable, we anchor the study on the bicycle architecture: the BB-based modular architecture from Tour de Gross~\cite{yoder_tour_2025}, shown in \Cref{fig:modular_arch}.
We use its line topology with one factory as the \emph{reference BB system} for the main end-to-end evaluation, since it is the smallest modular BB setting that already exposes both nearest neighbor communication constraints and shared factory constraints.
The one-factory assumption reflects a resource-constrained early regime where factory capacity is scarce~\cite{gidney_how_2025, gidney_magic_2024}, while richer qLDPC module connectivity or additional factory service would require extra adapter or extractor machinery~\cite{cross_improved_2024, williamson_low-overhead_2024, swaroop_universal_2024}.

On this reference system, 
we investigate the robustness of our compiler optimizations to a wider regime of BB-based systems by
sweeping the instruction cost ratios, logical processing unit count, and factory count.
Our workload suite contains more than 40 benchmark categories drawn from PennyLane~\cite{bergholm_pennylane_2022} and MQTBench~\cite{quetschlich_mqt_2023}, including quantum algorithms and Hamiltonian simulations with varying sizes.
The compiler optimizations are summarized in \Cref{tab:opt_list}.

Our main empirical result shows that in the studied regime, synthesizing arbitrary-angle rotations at the factory (syn@fac) rather than at the logical processing unit yields the largest reduction in estimated circuit failure probability: 9.0$\times$ on average across non-Clifford benchmarks.
The later instruction-level cost ratio, LPU count, and factory count sweeps preserve this qualitative ordering over the regimes we report.
This suggests that synthesis placement is the dominant bottleneck in this studied regime.

Besides assessing the bottleneck, we also develop a complete compilation pipeline for the studied systems, which is missing in Tour de Gross.
Transvection-based Clifford deferral is a key optimization for compiler practicality: it reduces Clifford deferral runtime by 77\% on the evaluated workloads, substantially improving compilation scalability for large-depth circuits.
Clifford insertion provides additional but secondary duration gains.

The rest of the paper proceeds as follows.
\Cref{sec:generalized_compilation_framework} introduces the compilation pipeline.
\Cref{sec:syn_at_fac,sec:transvection,sec:depth_opt} detail the three compiler contributions, and \Cref{sec:evaluation} present the evaluation.

% \begin{table}[ht]
% \centering
% \resizebox{\columnwidth}{!}{
% \begin{tabular}{lll}
% \toprule
% Name & Tour de Gross & This work \\
% \midrule
% Focus & ISA, primitives & pipeline and optimization \\
% Benchmarks & random circuit, TFIM & 14+ categories \\
% Qubit allocation & fixed & formalize and optimize \\
% Clifford deferral & Explicitly neglected & linear to circuit depth \\
% Clifford optimization & None & duration reduction \\
% \bottomrule
% \end{tabular}
% }
% \caption{
% Comparison between Tour de Gross and this work.
% }
% \label{tab:tour_vs_our}
% \end{table}

\section{Background}

\begin{figure*}[ht]
    \centering
    \includegraphics[trim={0 0.3cm 0 0}, width=0.94\linewidth]{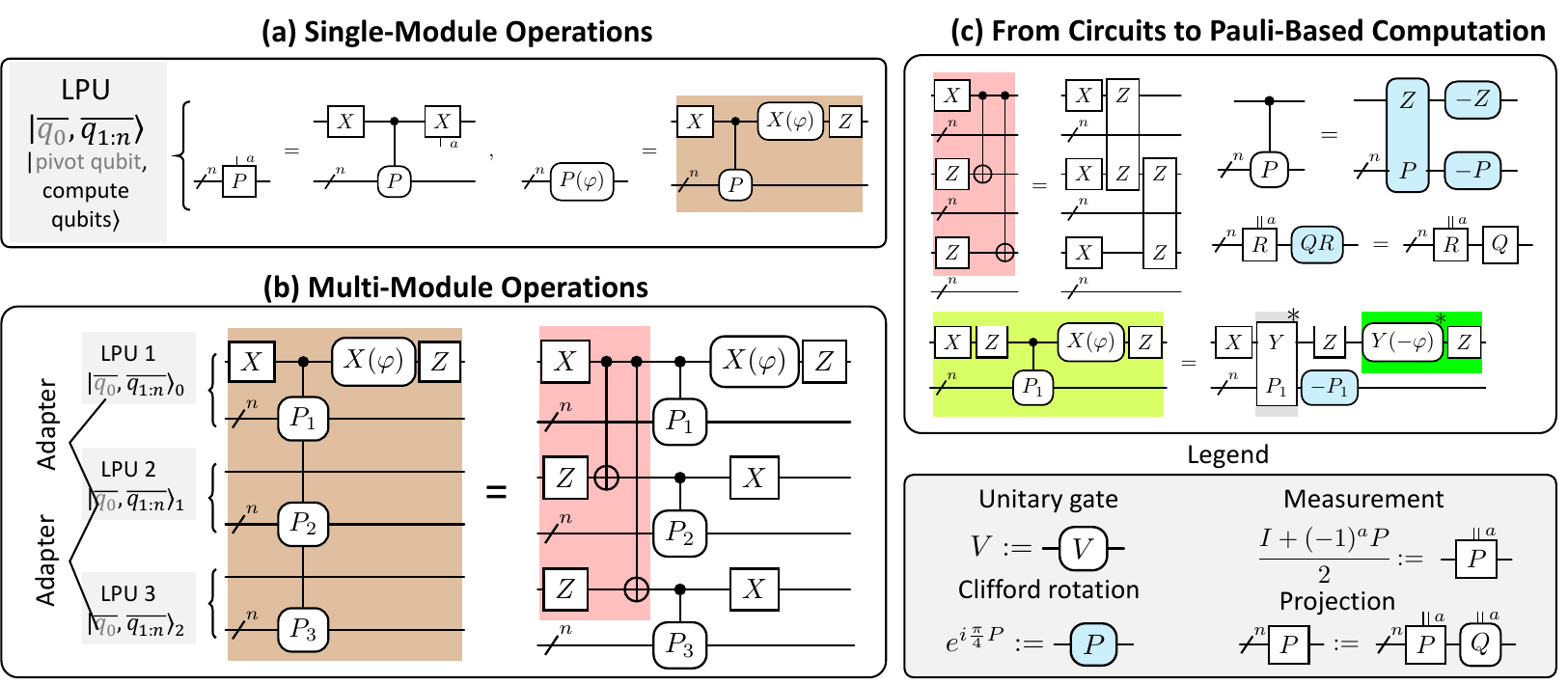}
    \caption{
    Lowering single-module and multi-module PBC operations to BB instructions.
    Two components highlighted in the same color are equivalent.
    (a) Single-module operations on an LPU.
    (b) Multi-module operations on three LPUs.
    (c) Gadgets that reduce multi-module operations to inter-module $ZZ$ measurements and in-module Pauli operations; the resulting lowered circuit is represented in \Cref{fig:pipeline}(d).
    }
    % \Description{}
    \label{fig:modular_compile}
\end{figure*}

\subsection{Logical Memory and the Bicycle Architecture}

Stabilizer codes are characterized by parameters $[[n,k,d]]$:
$n$ physical qubits encode $k$ logical qubits at code distance $d$. The encoding rate $k/n$ measures the qubit overhead per logical qubit. For example,
surface codes are $[[n,1,\Theta(\sqrt{n})]]$ codes whose stabilizers act only on nearest-neighbor qubits, a highly desirable property for hardware with limited connectivity. However, encoding only one logical qubit per block means that the encoding rate $k/n \rightarrow 0$ vanishes as the code grows, imposing severe qubit overhead at scale.

Recent breakthroughs in quantum low-density parity check (qLDPC) codes, where each stabilizer acts on $O(1)$ (possibly non-adjacent) qubits, have dramatically improved the encoding rate
\cite{krishna_fault-tolerant_2021, panteleev_degenerate_2021, panteleev_quantum_2022, tillich_quantum_2014, kovalev_quantum_2013}.
Since each qLDPC code block encodes many logical qubits ($k > 1$), this naturally aligns with modular architectures where a module houses one or more code blocks.
Bivariate Bicycle (BB) codes~\cite{bravyi_high-threshold_2024} are a finite-size instantiation achieving roughly 10-fold lower qubit overhead than surface codes.
In this paper, we focus on the $[[144,12,12]]$ BB code (i.e., the gross code), detailed in \Cref{app:background}.

The \emph{bicycle architecture}, first proposed in~\cite{yoder_tour_2025}, is a fault tolerant quantum computing architecture based on the BB codes. In a bicycle architecture, each module implements an \emph{extractor} on a code block that enables a tailored set of in-module Pauli measurements, while code--code adapters and code--factory adapters support inter-module measurements and magic state teleportation, respectively (\Cref{fig:modular_arch}).
We use logical processing unit (LPU) to denote the bundle of one pivot qubit and $k-1=11$ compute qubits in a module.
The pivot qubit is the ancilla that mediates Pauli rotations and Pauli measurements, while the compute qubits carry the workload data.
Inside an LPU, BB codes support a small set of fast symmetry-based Cliffords called shift automorphisms~\cite{yoder_tour_2025}.
Arbitrary-angle rotations, however, still rely on factory-produced magic states.
In the studied regime, the inter-module measurements and magic states are  among the most scarce resources in the architecture.

\subsection{Pauli-Based Computation}
\label{sec:pbc}

Fault tolerant computation on the bicycle architecture is expressed in the Pauli-based computation (PBC) framework~\cite{litinski_game_2019}, where circuits are decomposed into Pauli rotations \(P(\theta)=e^{-i\theta P}\) and Pauli measurements.
This representation is natural for Clifford+$T$ circuits, and Clifford gates can commute through the circuit and deferred to the end, where they are absorbed into the final measurements.
After qubit allocation, each Pauli operation is either single-module, meaning its support lies within one LPU, or multi-module, meaning its support spans multiple LPUs.
%This distinction is the key interface between the compiler and the BB modular architecture.

\subsection{From PBC to native BB Instructions}
\label{sec:modular_compile}

\Cref{fig:modular_compile} summarizes the compilation rules from PBC circuits to BB instructions used throughout the paper, while \Cref{app:background} gives the full gadget-level derivation.

For \textbf{single-module operations}, such as a Pauli rotation or Pauli measurement, the pivot qubit of that LPU serves as the ancilla, and the entire operation is executed locally within the module, either by in-module measurement synthesis or Pauli rotation synthesis.

For \textbf{multi-module operations}, the Pauli operator is decomposed into factors \(P_1 \otimes P_2 \otimes\dots\).
The pivot qubits of the participating LPUs are first coordinated through the code--code adapters. Each module then applies its local controlled-$P_i$ piece.
Multi-module Pauli measurements follow the same structure, except the first pivot is measured in the Pauli-$X$ basis instead of driving a rotation.
The resulting construction is then reduced to BB instructions consisting of inter-module $ZZ$ measurements together with in-module Pauli operations, as shown in \Cref{fig:modular_compile}(c).
Consequently, once qubit allocation determines which operations are multi-module, the support of those operations directly determines the communication-induced inter-module measurements in the lowered circuit.

A separate source of inter-module measurements comes from delivering magic states from the factory to an LPU for arbitrary-angle rotations.
We refer to those as teleportation-induced inter-module measurements, which constitute the dominant inter-module communication cost for workloads with many arbitrary-angle rotations.

\subsection{In-Module Measurement Synthesis}
\label{sec:bb_meas}

For the gross code, practical extractors expose a small generating set of logical Pauli measurements~\cite{yoder_tour_2025}.
Shift automorphisms enlarge that set into \emph{native measurements}, and the same primitives provide \emph{native Clifford rotations}.
These native Clifford rotations generate the module-level Clifford group, so arbitrary in-module Pauli measurements on the 11 compute qubits are synthesizable.

What matters for this paper is the resulting cost landscape.
The number of native measurements required to synthesize an in-module Pauli measurement is highly nonuniform, with cost levels of 1, 7, 13, 19, and 25 native measurements.
This asymmetry later enables Clifford insertion to shorten the circuit duration.

\subsection{Related Work}

Tour de Gross~\cite{yoder_tour_2025} is the direct baseline for this paper: it defined the BB modular architecture and instruction set.
However, it does not provide a complete compilation pipeline, and it evaluates only two benchmarks: random circuit and transverse-field Ising model.
This paper constructs a complete compilation pipeline and asks a different architecture question, namely which compiler optimizations most reduce estimated circuit failure probability.
Other modular FTQC work has focused more on architecture construction, entanglement generation, hardware scaling, defects, or catastrophic-error handling than on end-to-end compiler bottlenecks for BB code computation~\cite{haug_lattice_2025, kim_fault-tolerant_2024, lin_codesign_2024, singh_modular_2024, wu_mitigating_2025, xu_distributed_2022}.

Most PBC compilers in the literature target monolithic or surface code settings~\cite{herzog_lattice_2025, hirano_locality-aware_2025, kan_sparo_2025, kobori_lsqca_2024, wang_optimizing_2024, wang_tableau-based_2025}, which is outside the scope of this paper.
Same-time work by Sethi \emph{et al.}~\cite{sethi_optimizing_2026} studies communication-aware qubit allocation for BB code architectures, which we adopt as a backend pass rather than a contribution.
Our evaluation suggests that qubit allocation has up to 5.4\% end-to-end estimated circuit failure probability reduction on the studied regimes and benchmarks, and thus our main conclusion is unaffected.
Gate-based distributed compilers optimize SWAPs and two-qubit communication~\cite{du_hardware-aware_2025, jeng_modular_2025, mengoni_efficient_2025, sang_toward_2025, wu_autocomm_2022, wu_collcomm_2022, wu_qucomm_2023}; our setting instead centers on high-weight Pauli operations and arbitrary-angle rotation synthesis, so the dominant compiler techniques and tradeoffs differ.

\section{Compilation Pipeline}
\label{sec:generalized_compilation_framework}

\begin{figure}[t]
    \centering
    \includegraphics[trim={0 0.4cm 0 0}, width=0.9\linewidth]{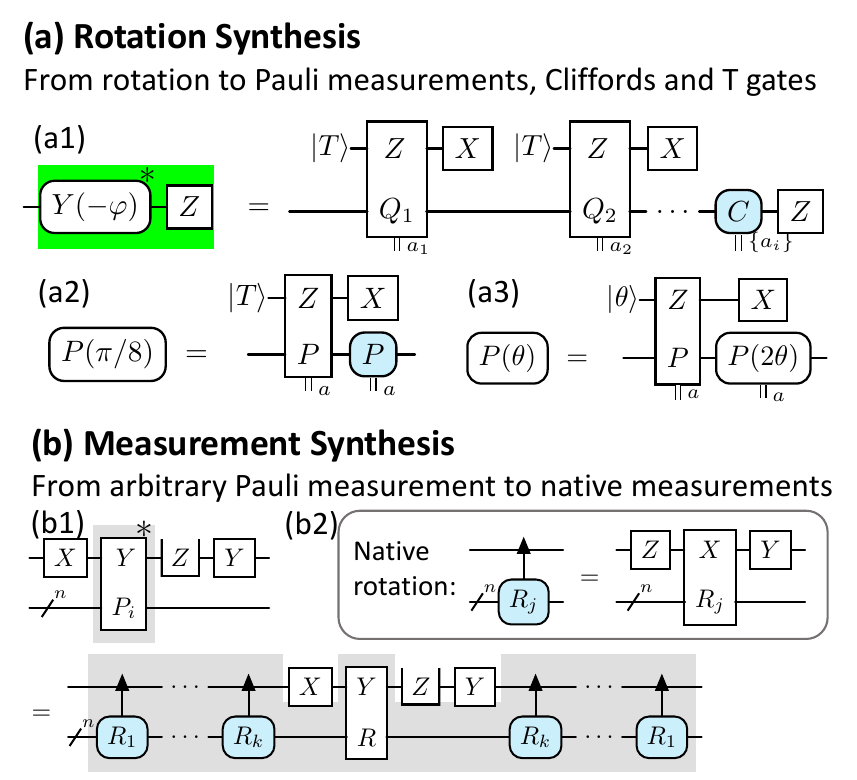}
    \caption{
    (a) Synthesize arbitrary-angle rotation by gate teleportation under the two placement choices
    (\Cref{sec:small_angle_synthesis,sec:syn_at_fac}).
    (b) Synthesize arbitrary Pauli measurement by native measurements and native rotations
    (\Cref{sec:bb_meas,sec:depth_opt}).
    }
    % \Description{}
    \label{fig:meas_and_rot_synth}
\end{figure}

\subsection{Pipeline Overview}

This section presents the compilation pipeline shown in \Cref{fig:pipeline}.
We begin with a Pauli-based computation circuit that contains Clifford rotations, arbitrary-angle rotations, and measurements.
First, we defer Cliffords to the end and absorb them into later measurements~\cite{litinski_game_2019}, leaving only arbitrary-angle rotations and Pauli measurements.
Next, we assign logical qubits to modules using the reproduced allocator of Sethi \emph{et al.}~\cite{sethi_optimizing_2026}.
The allocation determines whether each operation is single-module or multi-module, after which we apply the PBC-to-BB transformation from \Cref{sec:modular_compile} to obtain BB instructions.
Finally, we build a weighted directed acyclic graph (DAG) of the transformed circuit for duration analysis and later critical-path optimization.

For evaluation, we separately count teleportation-induced inter-module measurements from arbitrary-angle rotation synthesis and communication-induced inter-module measurements from multi-module support after allocation.
This separation lets us identify which part of the pipeline dominates estimated circuit failure probability in the studied regime.

\subsection{Rotation Synthesis Placement}
\label{sec:small_angle_synthesis}

For arbitrary-angle rotations, we use Ross-Selinger synthesis~\cite{ross_optimal_2016} implemented in \verb|pygridsynth|~\cite{selinger_pygridsynth_2018}.
Given a target \(P(\varphi)\), the decomposition produces a sequence of \(\pi/8\) rotations followed by a trailing Clifford, which we again defer into later measurements.
In this paper, the main compiler decision is how this synthesis is placed in the modular architecture.

We synthesize arbitrary-angle rotations \emph{after} qubit allocation rather than before.
Early synthesis tends to deepen the intermediate circuit and enlarge Pauli support once by-product Cliffords are deferred, which hurts both compilation runtime and the later estimated circuit failure probability under the studied cost model.

After allocation, the central remaining choice is \emph{where} synthesis is finished.
\textbf{syn@LPU} teleports \(\ket{T}\) states to the LPU and completes synthesis locally (\Cref{fig:meas_and_rot_synth}(a1)), whereas \textbf{syn@fac} synthesizes \(\ket{\theta},\ket{2\theta},\dots\) at the factory and teleports those states instead (\Cref{fig:meas_and_rot_synth}(a3)).
Detailed analysis of estimated circuit failure probability and duration for this placement decision appears in \Cref{sec:syn_at_fac}.

\subsection{Remaining Optimizations}

The remaining optimizations intervene at two different stages of the pipeline.
Before allocation, the compiler still operates on a PBC circuit, and this is where transvection accelerates Clifford deferral for long Clifford-rich workloads (\Cref{sec:transvection}).
After allocation and transformation, the circuit is represented as BB instructions with explicit instruction latencies and dependencies, enabling weighted DAG analysis.
At that stage, in-module Pauli measurements inherit the nonuniform synthesis costs summarized in \Cref{sec:bb_meas}, while inter-module $ZZ$ measurements remain as separate communication operations.
This transformed representation is what later enables Clifford insertion to shorten the critical path (\Cref{sec:depth_opt}).

\section{Synthesize Arbitrary-Angle Rotations at Factory}
\label{sec:syn_at_fac}

For the studied regime,
inter-module communication is a primary contributor
to estimated circuit failure probability under the studied cost model.
However, the synthesis placement in Tour de Gross leads to large number of inter-module measurements introduced by magic state teleportation.
This section targets at reducing such inter-module measurements.

Under the current instruction logical error rates (\Cref{tab:bb_instr_cost}), inter-module measurements are far more
error-prone than in-module measurements and also substantially more error-prone
than prepared magic states.
We compare two strategies for arbitrary-angle rotation synthesis and show that
performing synthesis at the factory (syn@fac) reduces inter-module measurements introduced by magic state teleportation.
Essentially, we trade expensive inter-module teleportation for cheaper in-module computation, and thus lower the estimated circuit failure probability.

Throughout this paper, we assume a \emph{basic $\ket{T}$ factory model}:
(i) A factory produces one \(|T\rangle\) state at a time with production latency \(t_{\mathrm{T}}\) and logical error rate \(p_{\mathrm{T}}\), precluding parallel \(|T\rangle\) preparation and latency hiding via concurrency.
(ii) The factory exposes a single-item buffer, meaning at most one \(|T\rangle\) can be cached; the next state can be prepared while the current one is consumed, but no deeper prefetching is possible.
This model is intentionally simple and reflects the earliest regime we study, where qubit count and factory capacity are both limited.

Denote $\theta$ as the angle of a target arbitrary-angle rotation \( P(\theta) \) and $\theta_{\text{syn}}$ as the synthesized angle produced by Ross and Selinger's algorithm.
Let $n_{\text{T}}(\theta)$ be the number of $\ket{T}$ states needed to synthesize $P(\theta)$ with approximation error $\epsilon$, i.e., $|\theta - \theta_{\text{syn}}| \leq \epsilon$.
% In this work, we set $\epsilon = 10^{-3}$.

We consider two strategies for synthesizing arbitrary-angle rotations via gate
teleportation, collectively referred to as \emph{gate-teleported rotations}.

\begin{figure}[t]
    \centering
    \includegraphics[trim={0 0.5cm 0 0}, width=\linewidth]{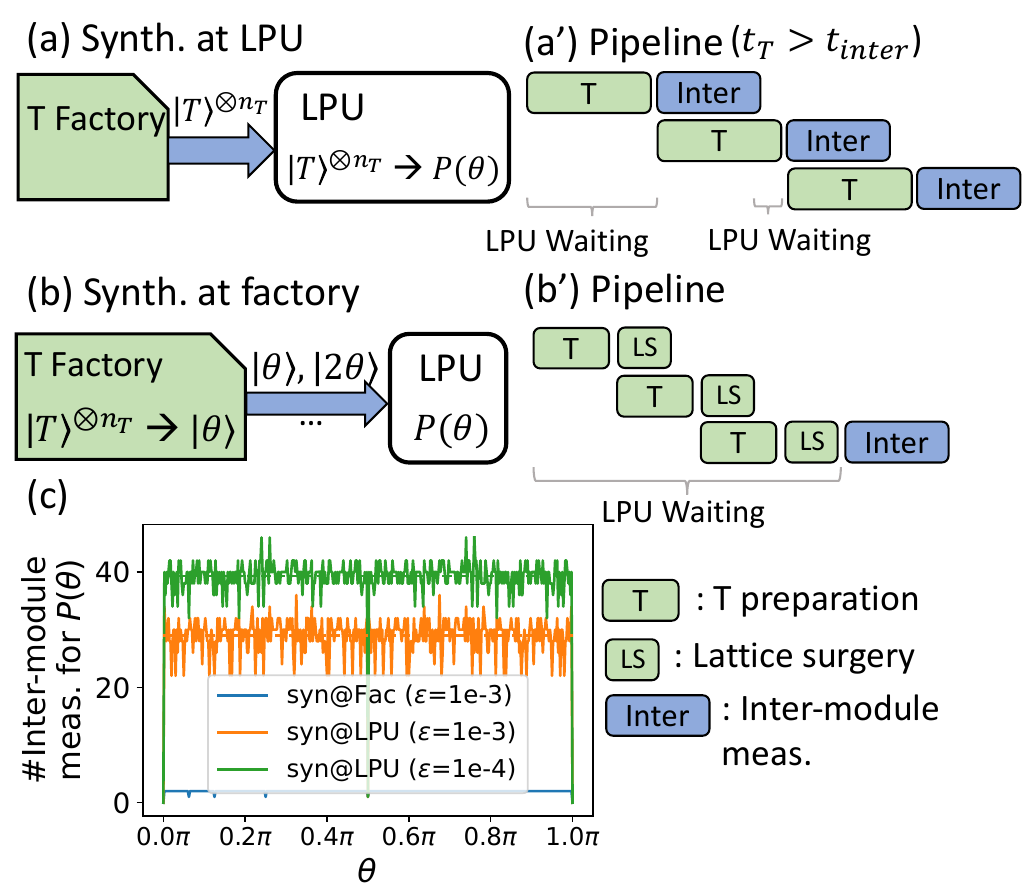}
    \caption{
    Synthesize an arbitrary-angle rotation $P(\theta)$ at the (a) LPU and (b) factory.
    (c) The number of inter-module measurements introduced by magic state teleportation for syn@fac and syn@LPU.
    After excluding points where the synthesis is trivial, i.e., $\theta \in\{ 0, \pi/2, \pi\}$,
    syn@LPU w/ $\epsilon = 10^{-3}$ has $\mu = 28.63$, $\sigma^2 = 7.57$; 
    syn@LPU w/ $\epsilon = 10^{-4}$ has $\mu = 39.25$, $\sigma^2 = 4.65$.
    }
    % \Description{}
    \label{fig:syn_at_fac_vs_lpu}
\end{figure}

The first strategy, \textbf{syn@LPU}, 
teleports $\ket{T}$ from the factory to the LPU and 
synthesizes \( P(\theta) \) at the LPU.
The diagram in \Cref{fig:syn_at_fac_vs_lpu}(a) shows each arrow from a factory to an LPU as a teleportation, and each teleportation requires one magic state teleportation, i.e., one inter-module measurement.
The circuit is shown in \Cref{fig:meas_and_rot_synth}(a1).
The second strategy, \textbf{syn@fac}, synthesizes arbitrary-angle magic states $\ket{\theta},\ket{2\theta},\dots$ at the factory using $\ket{T}$ states, 
teleports the resulting states to the LPU, and then applies \( P(\theta) \).
The diagram appears in \Cref{fig:syn_at_fac_vs_lpu}(b) and the circuit is shown in \Cref{fig:meas_and_rot_synth}(a3).
Implementing $P(\theta)$ from a teleported \(\ket{\theta}\) state yields a magic state teleportation result of $Z\otimes P$ as 1
with probability $1/2$, which triggers a correction $P(2\theta)$ using a $\ket{2\theta}$ magic state; otherwise the synthesis completes.
The $P(2\theta)$ correction reuses the same teleportation primitive recursively with $\theta$ replaced by $2\theta$~\cite{campbell_efficient_2016}.

\textit{Teleportation Count.}
As shown in \Cref{fig:syn_at_fac_vs_lpu}(c), syn@fac performs two magic state teleportations in expectation,
whereas syn@LPU performs $n_{\text{T}}(\theta)$ teleports
which averages to about 30 to 40.
This reduction in magic state teleportations is the structural reason syn@fac is favored by the first-order union bound estimate of estimated circuit failure probability.

\textit{T count.}
Relative to syn@LPU, syn@fac doubles the T count because it teleports two magic states on average~\cite{campbell_efficient_2016}
and each state costs $\overline{n}_{\text{T}}$ T gates to synthesize.

\textit{Qubit Overhead.}
Compared to syn@LPU, syn@fac requires one additional ancilla surface code patch at the factory to synthesize $\ket{\theta}$.

\textit{Failure-Probability Contribution.}
To keep the comparison controlled, we follow the instruction-counting framework of Tour de Gross~\cite{yoder_tour_2025} and use the same first-order union bound estimation that later defines estimated circuit failure probability in the evaluation.
% We update the magic state factory parameters to the same state-of-the-art factory model~\cite{sahay_fold-transversal_2025} for both the baseline and our method, so the differences we observe are attributable to compiler choices rather than inconsistent factory assumptions.
For syn@LPU, each $\ket{T}$ consumes one teleportation, giving the per-rotation contribution \( p_{\text{syn@LPU}}(\theta) \approx (p_{\text{inter}} + p_{\text{T}}) n_{\text{T}}(\theta) \).
For syn@fac, observe that for a fixed $\epsilon$, $n_{\text{T}}(\theta)$ is concentrated around its mean $\overline{n}_{\text{T}}$ for $\theta \in [0, \pi)$ as shown in \Cref{fig:syn_at_fac_vs_lpu}(c), 
yielding the per-rotation contribution \( p_{\text{syn@fac}}(\theta) \) as
$$
\sum_{k\ge 0}^{\infty} 2^{-k} (p_{\text{inter}} + n_{\text{T}}(2^k\theta)(p_{\text{T}}+p_{\text{ls}}))
\approx 2 (p_{\text{inter}} + \overline{n}_{\text{T}} (p_{\text{T}}+p_{\text{ls}})).
$$
Then the same cost model favors syn@fac when
$$
p_{\text{syn@fac}} < p_{\text{syn@LPU}}
\iff
(n_{\text{T}}-2)p_{\text{inter}} > n_{\text{T}} p_{\text{T}} + 2 n_{\text{T}} p_{\text{ls}},
$$
or equivalently
$$
\frac{p_{\text{T}}}{p_{\text{inter}}} + 2\frac{p_{\text{ls}}}{p_{\text{inter}}} < 1-\frac{2}{n_{\text{T}}}.
$$
In the present regime, nontrivial arbitrary-angle syntheses have representative $n_{\text{T}}$ values in the tens, as shown in \Cref{fig:syn_at_fac_vs_lpu}(c), so the right-hand side is already close to 1.
Thus the preference for syn@fac depends on the ratios on the left-hand side, in which $p_{\text{T}}/p_{\text{inter}}$ is swept in \Cref{sec:evaluation} to validate that our contribution is not an artifact of one exact instruction table.

\textit{Duration.}
Both strategies use the same Ross-Selinger decomposition and gate-teleportation primitive from \Cref{sec:small_angle_synthesis}, i.e.,
the same circuit in \Cref{fig:meas_and_rot_synth}(a1).
While consuming $\ket{T}$ via \( Z\otimes Q_{i+1} \) and $X$ measurements, we can prepare the next $\ket{T}$ in parallel, as illustrated in \Cref{fig:syn_at_fac_vs_lpu}(a',b').
We initiate synthesis as soon as an arbitrary-angle rotation is encountered.
With tighter scheduling and synchronization we could hide more latency by starting synthesis earlier, which we leave to future work.
For syn@LPU, \( t_{\text{syn@LPU}}(\theta) \approx n_{\text{T}}(\theta) \max(t_{\text{inter}}, t_{\text{T}}) + \min(t_{\text{inter}}, t_{\text{T}}) \),
and the time that LPU is waiting for T is \( t_{\text{T}} \).
Generally \( t_{\text{ls}} < t_{\text{T}} \), so \( t_{\text{syn@fac}}(\theta) \approx 2 (\overline{n}_{\text{T}} t_{\text{T}} + t_{\text{ls}} + t_{\text{inter}}) \),
and LPU waiting time is \( 2(\overline{n}_{\text{T}}t_{\text{T}} + t_{\text{ls}}) \).
Thus syn@fac can increase duration while sharply reducing estimated
circuit failure probability.
In \Cref{sec:evaluation}, we show that such increase due to idling has negligible effect on estimated circuit failure probability in the studies regime because teleportation
errors, not idling errors, dominate the execution bottleneck.
We further show that if there are two factories,
the end-to-end circuit duration of syn@fac is reduced to 1.05 times of syn@LPU,
almost completely compensating for the duration increase.

% Whether to use syn@LPU or syn@fac depends on specific hardware settings,
% which will be discussed in \Cref{sec:evaluation}.

% we replace the geometric weights by \( \mu_j=2^{-\ell - j} \) over levels $\ell$.
% we can approximate the error rate as
% \[
% \epsilon_{\text{synth@fac}}(\theta) \approx 2 \left[P_{\text{inter}} + \overline{n}_{\text{T}} \cdot P_{\text{T}}\right]
% \]
% If the angle is dyadic, i.e., $\theta = \pi/2^\ell$ for some $\ell\ge 0$
% known as Clifford level, its logical error rate is smaller 
% than $\epsilon_{\text{syn@fac}}(\theta)$ since we have early cut off of the arbitrary-angle correction.

\section{Efficient Clifford Deferral by Transvection}
\label{sec:transvection}

In this section, we argue that Clifford deferral (i.e. commuting Cliffords to the end of the
circuit and absorbed into final Pauli measurements~\cite{litinski_game_2019,wang_tableau-based_2025}) at \emph{compile
time} is nontrivial for realistic Clifford-rich and large-depth instruction streams.
In Tour de Gross, Clifford deferral is explicitly ignored in its compilation.

% In PBC, Clifford operations are cheap
% \emph{at execution time} because they can be commuted to the end of the
% circuit and absorbed into later Pauli measurements~\cite{litinski_game_2019,wang_tableau-based_2025}.
% For this reason, Tour de Gross explicitly says that it ignores this part of compilation,
% while we argue that carrying out this Clifford deferral efficiently at \emph{compile
% time} is nontrivial for realistic Clifford-rich and large-depth instruction streams.
% This section addresses compiler practicality rather than circuit failure
% probability directly.

For a PBC circuit with operation count $L$ on $n$ compute qubits,
the conventional right-to-left sweep updates all later Pauli rotations and
measurements whenever it encounters a Clifford~\cite{litinski_game_2019,wang_tableau-based_2025}.
Assuming each such Pauli update takes $O(n)$ time, this approach has
worst-case time complexity $O(nL^2)$, and in our benchmarks it becomes
prohibitively slow.

To address this bottleneck, we use \emph{transvections} as a compiler
primitive.
Prior transvection-based work has used symplectic transvections for logical
Clifford synthesis over stabilizer codes~\cite{rengaswamy_synthesis_2018} and
for fault-tolerant simulation via structured logical Trotter
circuits~\cite{chen_fault_2025}, while related work such as Pauli tracking
focuses on commuting Pauli operators through Clifford
circuits~\cite{ruh_quantum_2024}.
Our goal is different:
to accelerate Clifford deferral in an existing long PBC instruction stream.
The key idea is to scan from left to right, accumulate the effect of removed
Cliffords in a running symplectic transform, and apply that accumulated action
only when updating later Pauli rotations and measurements~\cite{dehaene_clifford_2003,koenig_how_2014}.

For each Pauli string $P$, we represent it by a symplectic vector $v = [v_x \mid v_z]$ of length $2n$,
where $v_x,v_z\in\mathbb{F}_2^n$.
$(v_x)_i = 1$ iff $X$ is on qubit $i$, and $(v_z)_i = 1$ iff $Z$ is on qubit $i$.
We denote the Pauli string corresponding to $v$ as $P_v$.
Then $P_v\,P_u = P_{v + u}$ up to phase.
We later abuse notation by identifying $P_v$ with $v$.
We define the \textit{symplectic inner product}
$
\langle v,\,u\rangle \;:=\; v_x\!\cdot u_z \;+\; v_z\!\cdot u_x \quad(\bmod\,2).
$
$P_v$ and $P_u$ anticommute iff $\langle v,u\rangle=1$, and commute otherwise.
For each nonzero $v$, the \textit{symplectic transvection} $T_v$ is the linear map
$
T_v(u)\;:=\;u\;+\;\langle v,u\rangle\,v \quad(\bmod\,2).
$
This map indicates that if $P_u$ and $P_v$ anticommute, i.e., $\langle v,u\rangle=1$, we update $P_u$ by multiplying
$P_v$, and do nothing otherwise.
When a Clifford $P_v(\pm \pi/4)$ conjugates an arbitrary Pauli rotation $P_u(\theta)$
or measurement $M_{P_u}$,
$u$ is updated to $T_v(u)$ up to phase.
The matrix form of the linear map $T_v$ is (abusing notation between the linear map and its matrix)
$$
T_v\; := \; I_{2n} + v (\Omega v)^\top,\quad 
\Omega\; :=\; \begin{bmatrix}
0 & I_n \\
I_n & 0
\end{bmatrix}.
$$

If the sequence of rotations or measurements is $v_0,v_1,\dots,v_{L-1}$ and the set of Clifford indices is
$C\subset\{0,\dots,L-1\}$, then for any non-Clifford operation or measurement $v_j$, it is updated as
$$
v_j \leftarrow \Big(\prod_{i\in C,\ i<j} T_{v_i}\Big)\, v_j,
$$
This product of $T_{v_i}$'s represents the accumulated effect of Cliffords sitting at the left of $v_j$.
The detailed steps are: (a) Initialize $S=I_{2n}$ and scan from left to right.
(b) When we encounter a Clifford $v$ to be removed, update $S \leftarrow S\,T_v$.
% $$
% S \leftarrow S\,T_v \;=\; S + (S\, v)\,(\Omega\, v)^\top
% $$
(c) When we encounter a non-Clifford rotation or measurement $v$, update $v \leftarrow S v$.

\emph{Complexity}: We traverse the sequence in one pass.
Updating $S$ once takes $O(n^2)$ time, so the total time complexity is
$O(n^2\cdot L)$.
This optimization is most beneficial when $L \gg n$, which is the regime that
arises for the realistic Clifford-rich workloads targeted in this paper.

\section{Duration Reduction by Clifford Insertion}
\label{sec:depth_opt}

In this section, we introduce Clifford insertion to reduce the duration of in-module Pauli measurement synthesis.
Unlike synthesis placement, this optimization is not the dominant driver of estimated circuit failure probability, but it yields useful critical-path improvements.

Prior surface code work inserts Cliffords statically to eliminate costly $Y$-basis measurements~\cite{stein_architectures_2024}.
However, this cannot be applied directly to BB codes since they have distinct Pauli measurement synthesis.
For BB codes, we propose a dynamic optimization tailored to the nonuniform cost of in-module Pauli measurement synthesis.

\begin{figure}[t]
    \centering
    \includegraphics[trim={0 0.5cm 0 0}, width=0.9\linewidth]{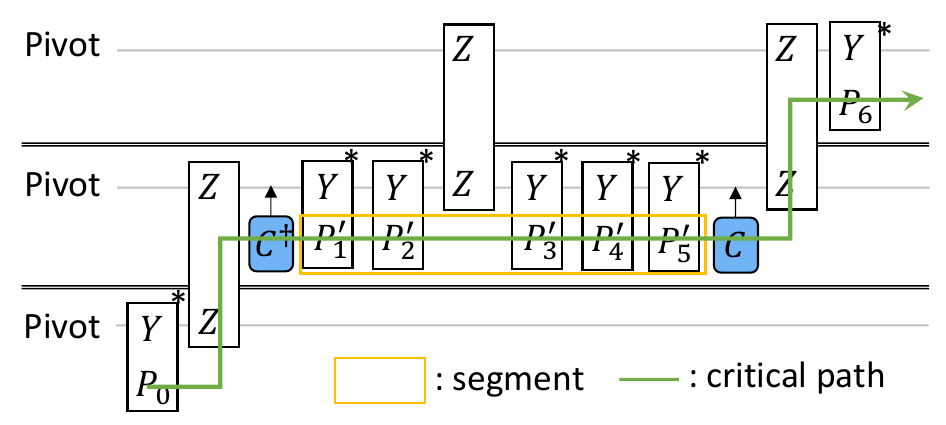}
    \caption{
        A segment is a consecutive sequence of in-module operations on the critical path.
        Clifford insertion optimizes the segment and does not affect pivot-only operations.
    }
    % \Description{}
    \label{fig:segment}
\end{figure}

After the PBC-to-BB transformation, we build a weighted directed acyclic graph (DAG) of BB instructions as in \Cref{fig:pipeline}(d).
We then find the critical path and partition it into \emph{segments}: maximal consecutive sequences of compute-qubit operations that stay within the same module, as shown in \Cref{fig:segment}.
A segment may span inter-module $ZZ$ measurements because those operations do not affect the optimization and their duration cannot be reduced.

For a segment $S = \{P_1, P_2, \dots, P_l\}$, let $f: \mathcal{P}_{n} \to \mathbb{R}^+$ denote the cost of synthesizing an in-module measurement.
We insert a Clifford $C^\dagger$ before the segment and $C$ after it, so that each $P_i$ becomes $C P_i C^\dagger$.
We seek Cliffords that reduce the total synthesis cost after paying for the inserted opening and closing Cliffords.
An illustrative objective is
\begin{equation*}
\max_{C} -f(C) - f(C^{\dagger}) + \sum_{i=1}^l \left(f(P_i) - f(C P_i C^{\dagger})\right).
\end{equation*}
For the BB codes studied here, $f$ takes values in $\set{1, 7, 13, 19, 25}$.

A single Clifford is often too rigid for a long segment, so we allow multiple disjoint \emph{windows}, i.e., $ C $ and $ C^\dagger $ pairs, within a segment.
For a fixed library of candidate Cliffords with cheap synthesis cost, we first score the raw reduction obtained at each position under conjugation and then use dynamic programming to choose the set of non-overlapping windows with maximum net reduction after charging the opening and closing overhead.
\Cref{sec:app_cliff_insertion} gives the full recurrence and backtracking details.

In practice, we apply this window-selection pass to each segment on the current critical path.
We then rank segments by expected net duration reduction, apply Clifford insertion to the top $T$ segments, rebuild the critical path, and iterate until the reduction relative to the previous iteration falls below $1\%$.
This keeps Clifford insertion localized to the part of the circuit.

\section{Evaluation}
\label{sec:evaluation}

\begin{figure*}[ht]
    \centering
    \resizebox{0.89\linewidth}{!}{  
        \includegraphics[trim={0 0.7cm 0 0},
            width=\linewidth]{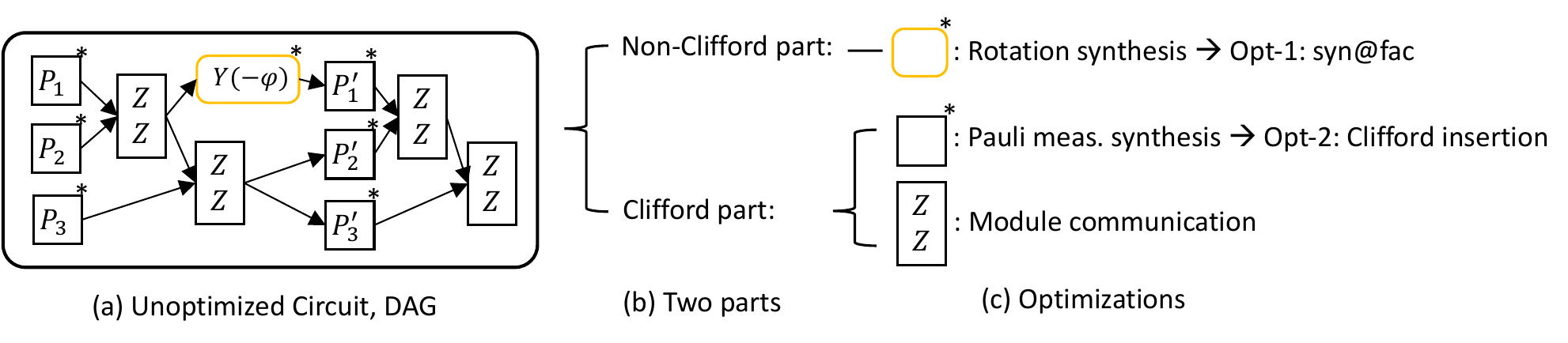}
    }
    \caption{
        Visualization of the adopted compilation optimizations.
        (a-b) A PBC circuit, in form of a DAG, can be viewed as two parts: the non-Clifford part and the Clifford part.
        (c) The non-Clifford part is arbitrary-angle rotation synthesis, which is optimized by syn@fac.
        The Clifford part consists of Pauli measurement synthesis and module communication.
        Clifford insertion optimizes the former, while module communication is handled using the reproduced allocator of Sethi \emph{et al.}~\cite{sethi_optimizing_2026}.
        Transvection accelerates Clifford deferral and is not shown because it affects compile time rather than circuit execution.
    }
    % \Description{}
    \label{fig:adopted_optimizations}
\end{figure*}

\begin{table}[ht]
    \centering
    \resizebox{0.85\columnwidth}{!}{
    \begin{tabular}{llll}
        \toprule
        % name & label($i$) & time($t_i$) & $p_i$, $p=1e-3$ & $p_i$, $p=1e-4$ \\
        Name & Label $i$ & Time $t_i$ & LER $p_i$ \\
        % & $i$ & $t_i$ & $p_i$ \\
        \midrule
        Idle & idle & 8 & $10^{-8.8}$ \\
        Shift automorphism & aut & 14 & $10^{-6.4}$ \\
        In-module meas. & in & 120 & $10^{-5.0}$ \\
        Inter-module meas. & inter & 120 & \boldmath{$10^{-2.7}$}\\
        Magic state teleport & tele & $t_{\text{inter}}$ & $p_{\text{inter}}$ \\
        % T prep., Gidney et al.~\cite{yoder_tour_2025,gidney_magic_2024} & T & 351 & $10^{-5.5}$ \\
        T prep.~\cite{sahay_fold-transversal_2025} & T & 122 & $2 \times 10^{-6}$ \\
        Lattice surgery~\cite{stein_architectures_2024} & ls & 66 & $10^{-7.2}$ \\
        \bottomrule\\
        \end{tabular}
    }
    \caption{Instruction table at the present regime, where physical error rate $p=10^{-3}$.
        Timesteps and logical error rates (LERs) of instructions are used to construct the cost model.
        Unit of $t_i$ is gate counts assuming each physical gate costs one timestep.
        Lattice surgery is charged only to syn@fac, i.e., absent in the baseline.
        We use the same magic state factory~\cite{sahay_fold-transversal_2025} for both our design and the baseline.
    }
    \label{tab:bb_instr_cost}
\end{table}

% \begin{figure}[ht]
%     \centering
%     \includegraphics[trim={0 0.7cm 0 0},
%    width=\linewidth]{figs/breakdown/mqtbench-m3-qchem-m2-gross1e3-sahay-0.0001.pdf}
%     \caption{
%         Breakdown of (a,b) circuit failure probability
%         and (c,d) circuit duration before and after all optimizations.
%         Clifford-only benchmarks (middle) and long circuits (right) are separated
%         because they operate on different scales.
%         Brown bars in (a,b) denote magic state preparation,
%         while brown bars in (c,d) denote LPU idling time waiting for magic states.
%     }
%     % \Description{}
%     \label{fig:breakdown_mqt_m3_qchem_m2_gross1e3}
% \end{figure}

% \begin{figure}[ht] % h: here, t: top, b: bottom, p: page
%     \centering
%     \includegraphics[trim={0 0.7cm 0 0},
%         width=\linewidth]{figs/breakdown/mqtbench-m3-3bar-gross1e3-sahay-0.0001.pdf}
%     \caption{
%         Ablation of duration.
%         Baseline: random partition, syn@LPU, Gidney et al.'s MSC, no Clifford insertion.
%         Baseline to middle bar: qubit allocation, syn@fac, Sahay et al.'s MSC w/ tuning.
%         Middle bar to right bar: Clifford insertion.
%     }
%     % \Description{}
%     \label{fig:breakdown_mqt_m3_3bar_gross1e3}
% \end{figure}

This section evaluates the compiler optimizations in the reference BB system under the present cost model, using estimated circuit failure probability as the primary metric.
Our main empirical result is that synthesizing arbitrary-angle rotations at the factory (syn@fac) gives the largest reduction in estimated circuit failure probability.
We then test the robustness of the syn@fac conclusion by sweeping instruction cost ratios, LPU count, and factory count.
We next show transvection-based Clifford deferral substantially reduces compile time, while Clifford insertion reduces duration.

Unless otherwise noted, all reported results (for both the optimized one and the baseline) use the reproduced allocator of same-time work by Sethi \emph{et al.}~\cite{sethi_optimizing_2026}, which we use as a backend pass rather than as a contribution of this paper.

\emph{Benchmark Information.}
Our workload suite contains
molecular Hamiltonian simulations in PennyLane~\cite{bergholm_pennylane_2022} (QChem) and
algorithms from MQTBench~\cite{quetschlich_mqt_2023}.
Each BB module carries 12 logical qubits:
11 compute qubits plus one pivot qubit.
Given $M$ modules we run every benchmark whose qubit count lies in $[11(M-1)+1, 11M]$.
QChem circuits are derived by Bravyi--Kitaev mapping~\cite{bravyi_fermionic_2002}
and Trotterization~\cite{suzuki_general_1991} using PennyLane's APIs with their default parameters.
We label each benchmark as (Category, \#Size).
For each category, we report the smallest, median, and largest instance (if available).

\emph{Estimated Circuit Failure Probability.}
We use the same first-order union bound as Tour de Gross~\cite{yoder_tour_2025}
to estimate circuit failure probability under the present cost model:
$
p_{\text{circ}} \approx \sum_{i \in I} N_i p_i,
$
where $I =\, $\{in, inter, idle, aut, tele, T, ls\},
$p_i$ is the logical error rate of each instruction,
and $N_i$ counts occurrences in the compiled circuit.
% This approximation is standard because simulating an entire fault-tolerant circuit is impractical.
Because full fault-tolerant circuit simulation is impractical, we use the present cost model as a proxy for identifying dominant bottlenecks in the studied regime and for comparing the relative effectiveness of our optimizations.
Accordingly, we separately report inter-module measurements induced by magic state teleportation and by module communication.
Some optimizations target only Clifford operations in the circuit, so we also report a Clifford-only metric that excludes gate-teleported rotations:
$
p_{\text{Cliff}} \approx
    \sum_{i \in I\setminus\{\text{tele}, \text{T}, \text{ls}\}} N_i p_i.
$

\emph{Circuit Duration.}
We build a DAG where nodes represent instructions
(\Cref{fig:pipeline}(e)).
Each node stores its instruction latency,
and the critical path is the longest weighted path.
Circuit duration, defined as $ t_{\text{circ}} $, equals the sum of latencies along that path.
To obtain Clifford-only duration we remove non-Clifford nodes
on the critical path.

\subsection{Overall Performance and Bottleneck Analysis}
\label{sec:overall_performance}

\begin{figure}[ht]
    \centering
    \resizebox{\linewidth}{!}{  
    \includegraphics[trim={0 0.7cm 0 0},
            width=\linewidth]{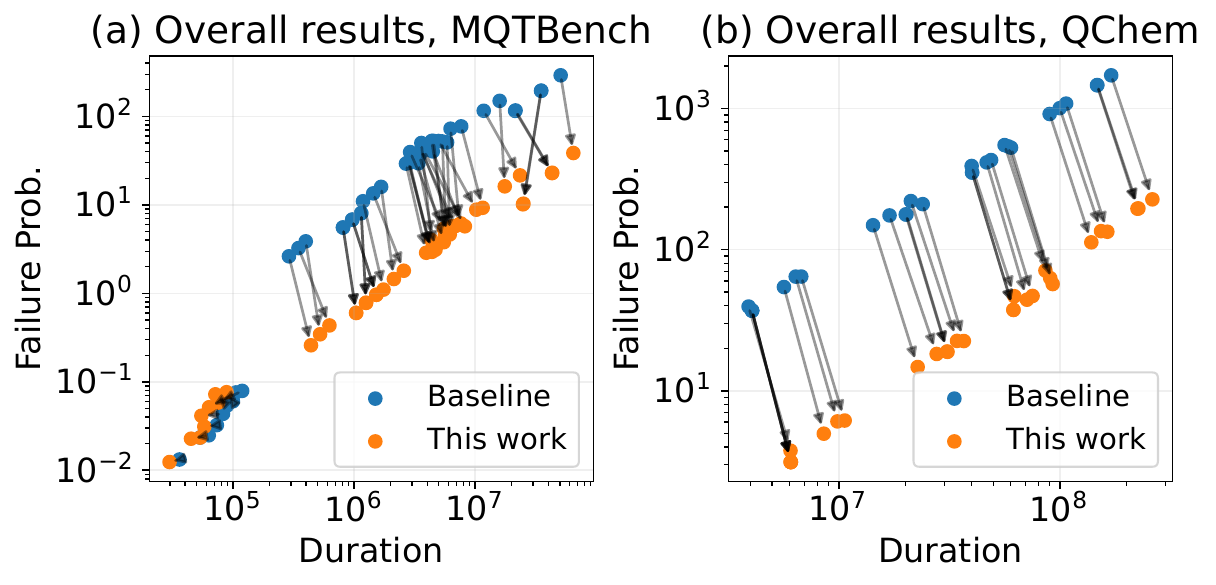}
    }
    \caption{
        Overall performance comparison between baseline and the compiler contributions of this work in \Cref{tab:opt_list} for the reference system.
    }
    % \Description{}
    \label{fig:baseline_vs_our}
\end{figure}

\begin{figure*}[ht]
    \centering
    \resizebox{\linewidth}{!}{  
    \includegraphics[trim={0 0.7cm 0 0},
            width=\linewidth]{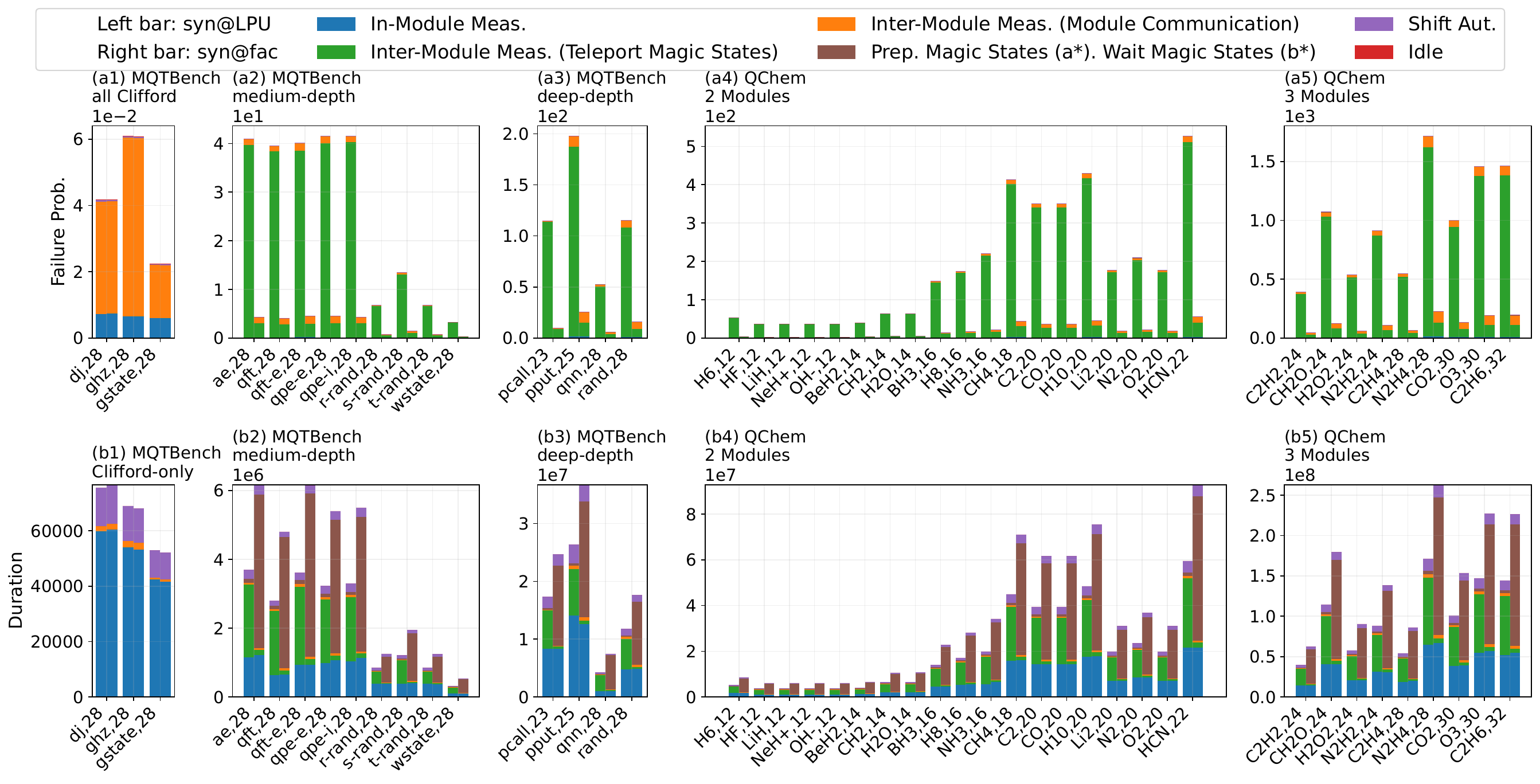}
    }
    \caption{
        Ablation: syn@fac vs the baseline (syn@LPU) for the reference system. Breakdown of failure probability and duration.
        We separate inter-module measurements into those introduced by magic state teleportation and those introduced by module communication.
        The figure shows that (i) syn@fac reduces the modeled failure probability substantially while increasing the duration. (ii) The increase in duration is mainly due to waiting for magic states (brown bars), which has little effect on estimated circuit failure probability because BB codes have low idling error rates. (iii) In the reference system, the inter-module measurements introduced by module communication are not the dominant contributor to the failure probability and duration.
    }
    \label{fig:breakdown_lpu_vs_fac}
    % \Description{}
\end{figure*}

We compare the baseline derived from Tour de Gross to our design where all the optimizations are adopted, as listed
in \Cref{tab:opt_list}.
The estimated circuit failure probability and duration are shown in
\Cref{fig:baseline_vs_our}, and their breakdowns are shown in \Cref{fig:breakdown_lpu_vs_fac}.

Within the reference system, our design achieves on average a 9.0$\times$ reduction in estimated circuit failure
probability across the non-Clifford benchmarks,
as shown in \Cref{fig:baseline_vs_our}(a,b).
% The reason is that syn@fac replaces roughly $n_{\text{T}}(\theta)$ magic state teleports per arbitrary-angle rotation with 2 in expectation.
For Clifford-only benchmarks, where magic state teleportation is absent, the gains are correspondingly smaller.

\textbf{Bottleneck: Magic State Teleportation.}
% Under the present cost model, the main contributor to the estimated circuit failure probability is
% inter-module measurements introduced by magic state teleportation.
% Under the adopted allocator, the remaining communication induced inter-module measurements are secondary in the reference system.
The estimated circuit failure probability of syn@LPU is dominated by inter-module measurements induced by magic state teleportation (green bars, \Cref{fig:breakdown_lpu_vs_fac}(a2-a5)).
syn@fac reduces such measurements from
$\overline{n}_{\text{T}}$ to 2 on average, which is the main reason it lowers
estimated circuit failure probability so sharply.
For Clifford-only circuits (\Cref{fig:breakdown_lpu_vs_fac}(a1)), no magic state teleportation is required and thus the changes are small.
Under the adopted allocator, the communication-induced inter-module measurements are secondary in the reference system (orange bars, \Cref{fig:breakdown_lpu_vs_fac}(a2-a5)).

The cost of syn@fac is increased circuit duration, mainly from LPU idling while waiting for factory-produced states (brown bars, \Cref{fig:breakdown_lpu_vs_fac}(b2-b5)).
Under the present cost model, this idling has little effect on estimated circuit failure probability because the idle logical error rate is much lower than the inter-module measurement error rate.
Thus, syn@fac trades fewer high-error teleportation-induced inter-module measurements for more low-error idling.
The factory-count sweep in \Cref{sec:sensitivity_of_syn_at_fac} shows that adding one more factory reduces the geometric-mean duration ratio from 1.5704 to 1.0474 and almost compensate for the duration increase.

\subsection{Scope and Sensitivity of syn@fac}
\label{sec:sensitivity_of_syn_at_fac}

The previous syn@fac analysis is established on the reference system.
We now test whether the syn@fac conclusion persists when varying the cost model, LPU count, and factory count.

\begin{figure}[ht]
    \centering
    \resizebox{\linewidth}{!}{  
        \includegraphics[trim={0 0.7cm 0 0},
            width=\linewidth]{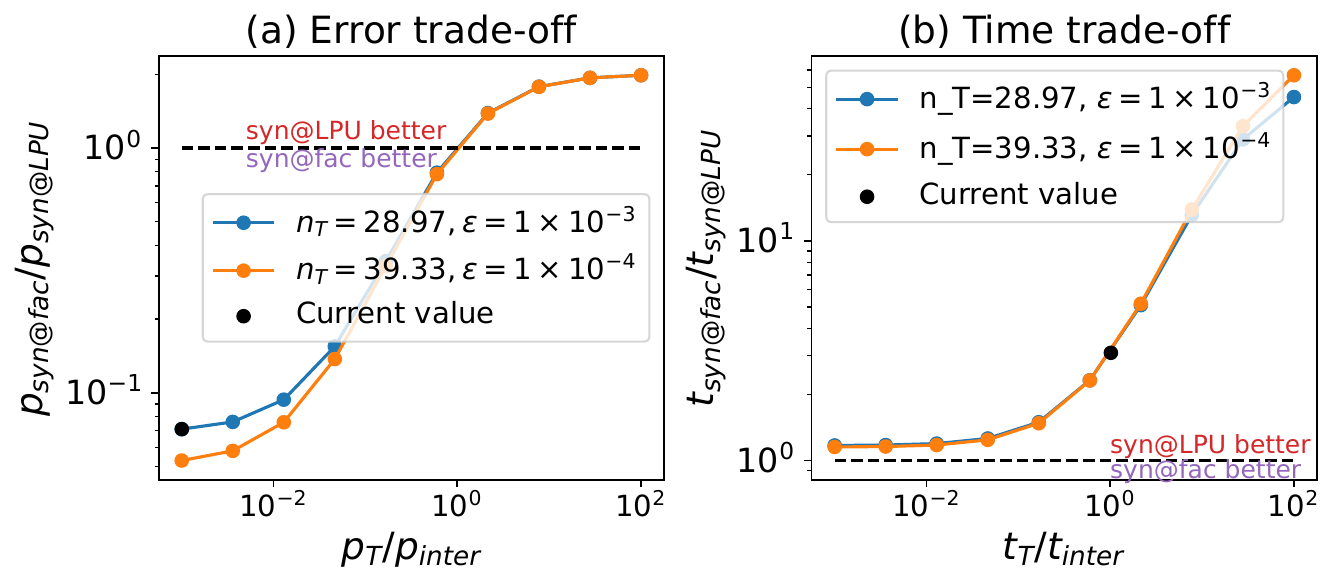}
    }
    \caption{
        Cost-ratio sensitivity of syn@LPU versus syn@fac for one arbitrary-angle rotation.
        Panel (a) varies the error-ratio terms that govern the first-order error comparison, and panel (b) varies the latency-ratio terms that govern the duration trade-off.
        The black dot marks the reference system used elsewhere in the paper.
    }
    % \Description{}
    \label{fig:cmp_syn_at_fac_vs_lpu}
\end{figure}

\textbf{Cost Model Ratio Sweep.}
We sweep the cost model ratios, $p_{\text{T}}/p_{\text{inter}}$ and $t_{\text{T}}/t_{\text{inter}}$, that most directly control the syn@fac versus syn@LPU comparison.
This ratio-based sensitivity analysis abstracts away most of the BB-specific absolute constants.
\Cref{fig:cmp_syn_at_fac_vs_lpu}(a) varies the error-ratio terms that govern the analytic bound in \Cref{sec:syn_at_fac}, while \Cref{fig:cmp_syn_at_fac_vs_lpu}(b) varies the latency-ratio terms that govern the duration trade-off.
The black dot marks the present cost model from \Cref{tab:bb_instr_cost}.
Note here the ratio sweep is intentionally a per-rotation evaluation but not an end-to-end one because its points are hypothetical instruction-ratio perturbations, not correspond to actual designs or implementations.

At the present cost model and \(\epsilon=10^{-3}\), syn@fac contributes about \(0.07\times\) the estimated failure probability of syn@LPU for one arbitrary-angle rotation.
This large gap follows from the teleportation count:
syn@LPU pays roughly \(n_{\text{T}}\) teleportation-induced inter-module measurements per arbitrary-angle rotation, while syn@fac pays 2 in expectation.
The stricter tolerance \(\epsilon=10^{-4}\) further favors syn@fac because it increases \(n_{\text{T}}\).
In contrast, syn@fac has higher per-rotation latency over the swept range under the basic factory model and scheduling assumption.
As mentioned in \Cref{sec:overall_performance}, such increased latency has negligible effect on end-to-end estimated circuit failure probability; besides, the end-to-end duration increase can be compensated.

% A faster factory might be able to mitigate the increase on duration,
% where the ratio of the duration of syn@fac to syn@LPU converges to 1.1.

% Together, panels (a) and (b) separate two facts:
% the estimated failure probability advantage of syn@fac is robust over a broad range of error-ratio assumptions, while its duration overhead is mainly sensitive to the factory-service ratio $ t_{\rm T}/t_{\rm inter} $.

\begin{figure}[ht]
    \centering
    \resizebox{\linewidth}{!}{
        \includegraphics[trim={0 0.7cm 0 0},
            width=\linewidth]{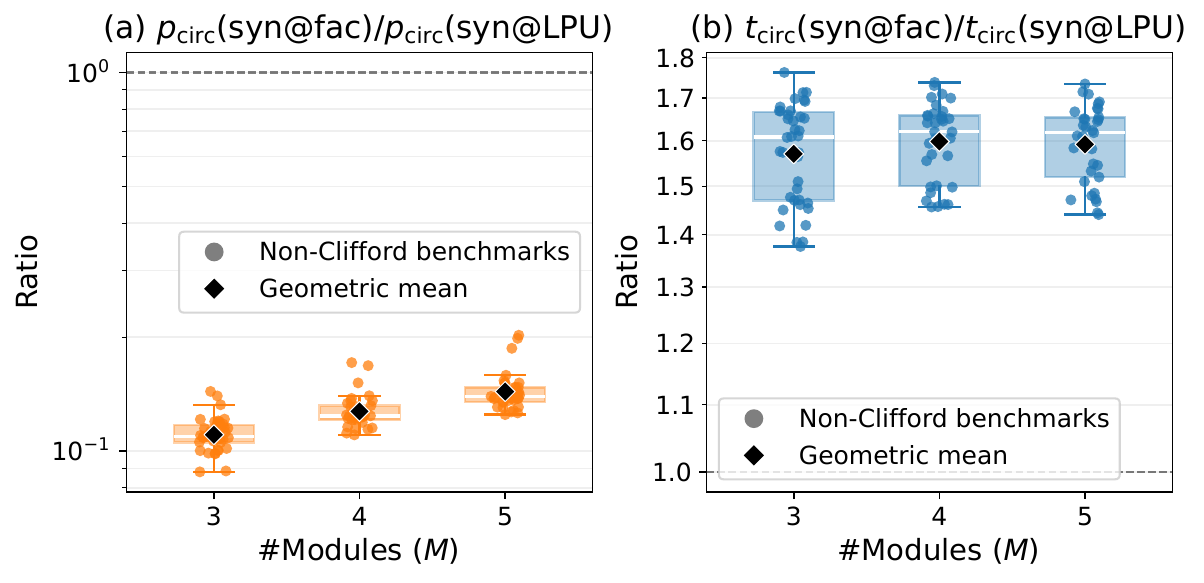}
    }
    \caption{
        LPU-count sensitivity around the reference system.
        We evaluate end-to-end on all non-Clifford benchmarks.
        Panel (a) plots the ratio of estimated circuit failure probability, $p_{\text{circ}}(\text{syn@fac})/p_{\text{circ}}(\text{syn@LPU})$, and panel (b) plots the ratio of circuit duration, $t_{\text{circ}}(\text{syn@fac})/t_{\text{circ}}(\text{syn@LPU})$.
        Across $M=3$--$5$, the failure-probability advantage of syn@fac is preserved, while its duration overhead remains nearly constant under a single shared factory.
    }
    % \Description{}
    \label{fig:sweep_num_lpu}
\end{figure}

\textbf{LPU Count Sweep.}
We fix line topology and one magic state factory, and include all non-Clifford benchmarks. We denote end-to-end estimated circuit failure probability as $p_{\text{circ}}$ and circuit duration as $t_{\text{circ}}$.
As shown in \Cref{fig:sweep_num_lpu}, the geometric mean of
$p_{\text{circ}}(\text{syn@fac})$ over $p_{\text{circ}}(\text{syn@LPU})$
increases only modestly from 0.111 at $M=3$ to 0.127 at $M=4$ and 0.144 at $M=5$.
That is, syn@fac still reduces estimated circuit failure probability by about a factor of 7.0 to 9.0 across this range.
By contrast, the geometric mean duration ratio
$t_{\text{circ}}(\text{syn@fac})$ over $t_{\text{circ}}(\text{syn@LPU})$
stays nearly flat at 1.57, 1.60, and 1.59 for $M=3$, $4$, and $5$, respectively.
This means that increasing the number of LPUs from 3 to 5 does not overturn the main conclusion:
under fixed line topology and one factory, syn@fac remains the stronger lever for reducing estimated circuit failure probability, although its advantage shrinks modestly as $M$ grows, indicating that module communication becomes relatively more important in larger systems.

\begin{figure}[ht]
    \centering
    \resizebox{\linewidth}{!}{
        \includegraphics[trim={0 0.7cm 0 0},
            width=\linewidth]{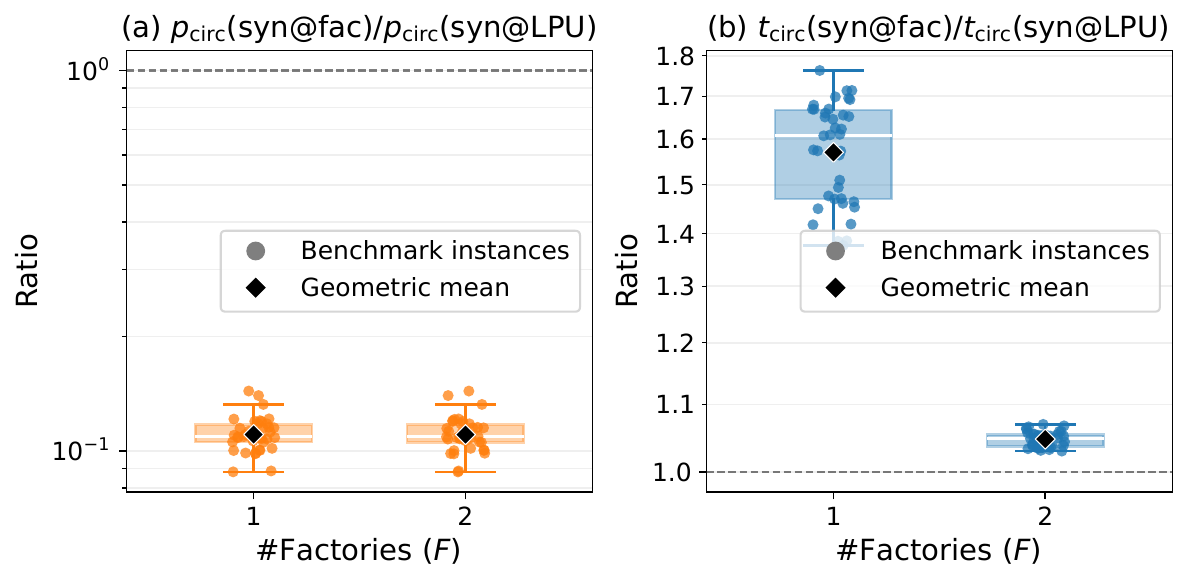}
    }
    \caption{
        Factory count sensitivity around the reference system under fixed line topology and the same reference cost model.
        Panel (a) plots the geometric mean of the estimated circuit failure-probability ratio, $p_{\text{circ}}(\text{syn@fac})/p_{\text{circ}}(\text{syn@LPU})$, for $F=1$ and $F=2$.
        Panel (b) plots the geometric mean of circuit-duration ratio, $t_{\text{circ}}(\text{syn@fac})/t_{\text{circ}}(\text{syn@LPU})$, for $F=1$ and $F=2$.
        Increasing $F$ leaves the error ratio nearly unchanged while sharply reducing the duration ratio.
    }
    % \Description{}
    \label{fig:sweep_num_fac}
\end{figure}

\textbf{Factory Count Sweep.}
We also vary factory count $F$, one concrete way to change factory service capacity, to identify the effect of the one-factory service constraint.
Increasing $F$ from 1 to 2 leaves the geometric-mean estimated failure probability ratio nearly unchanged, from 0.1105 to 0.1104,
while reducing the geometric-mean duration ratio from 1.5704 to 1.0474.
This shows that the estimated failure probability advantage of syn@fac is essentially unchanged, while most of its duration overhead comes from the one-factory service bottleneck.
The duration overhead is mainly due to LPU idling while waiting for magic states, so higher factory service can almost completely compensate for it, as discussed in \Cref{sec:overall_performance}.
However, we do not present $F>1$ as part of the reference design, because adding factories is not a free architectural change in this BB setting:
factories are expensive, extra factories require additional code--factory adapters, and supporting two factories per LPU may require stronger connectivity assumptions so that both factories can teleport magic states to the same pivot qubit on the same LPU (otherwise, extra conditions are needed for the correctness of teleportation in \Cref{fig:meas_and_rot_synth}(a)).

\subsection{Effectiveness of Transvection}
% The conventional method to defer Cliffords 
% is scanning the circuit from right to left;
% when encounter a Clifford,
% we update all rotations and measurements at its right.
% Our method is scanning the circuit from left to right,
% and using transvection to accumulate the effects of removed Cliffords
% and update remaining operations (\Cref{sec:transvection}).
We compare the conventional and our transvection-based Clifford deferral method
by runtimes on MacBookPro with M1 Pro chip, using Python and only one process,
and derive 77.04\% reduction for 3-module-size benchmarks
and 98.86\% reduction for 122-qubit benchmarks (\Cref{tab:eval_transvection}).
This result is separate from estimated circuit failure probability, since transvection improves compiler scalability but not circuit execution.

\begin{table}[ht]
    \centering
    \resizebox{\columnwidth}{!}{
    \begin{tabular}{llllll}
        \toprule
        Size Range & \#Circ. & Max \#Op. & Conventional & Ours & Reduction \\
        \midrule
        $[23, 33]$ & 159 & 41K & 123.03 s & 27.76 s & 77.04\% \\
        % $[122, 132]$ & 126 & 3 hrs (est.) & 5.13 min & 97\% (est.) \\
        $122$ & 14 & 161K & 46.43 min & 31.66 s & 98.86\% \\
        \bottomrule\\
        \end{tabular}
    }
    \caption{
    Comparison between conventional method and our transvection method, when deferring Clifford operations.}
    \label{tab:eval_transvection}
\end{table}

\subsection{Secondary Optimizations}

\begin{figure}[t] % h: here, t: top, b: bottom, p: page
    \centering
    \includegraphics[trim={0 0.7cm 0 0},
        width=\linewidth]{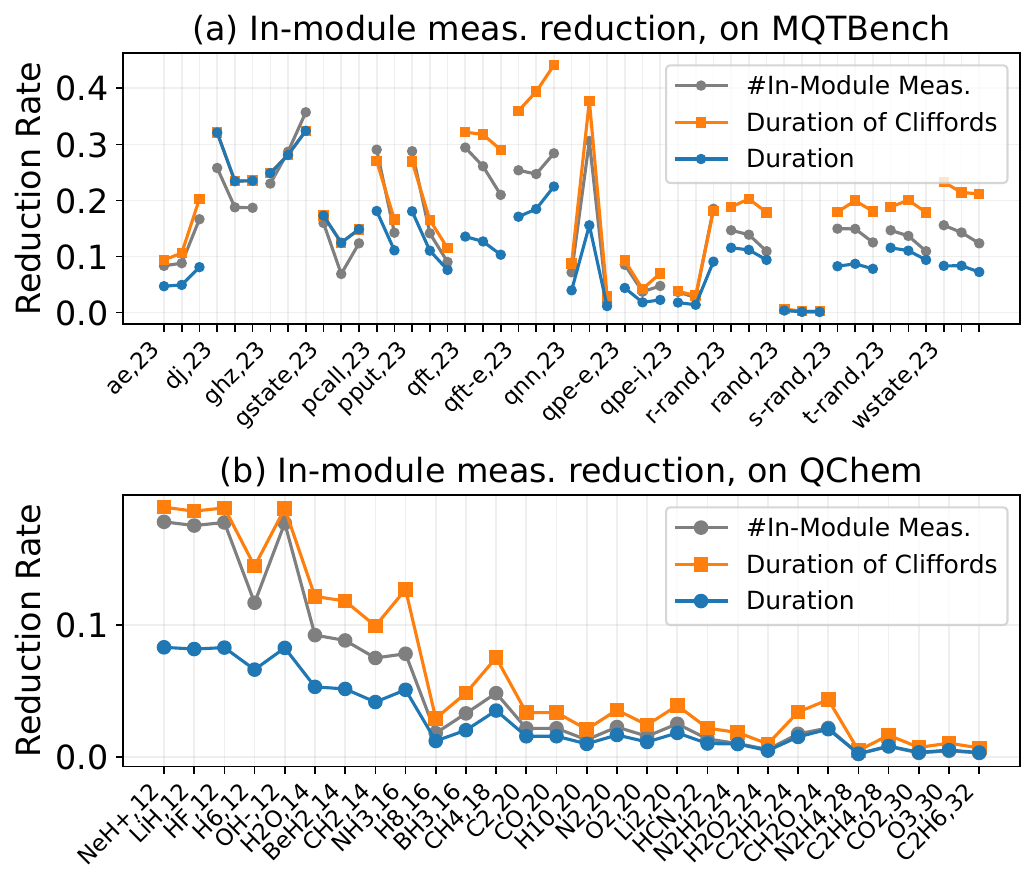}
    \caption{
        Ablation: only Clifford insertion is applied vs the baseline.
        For each category we plot the smallest, median, and largest instance
        (if available) and label the first instance on the x axis.
    }
    % \Description{}
    \label{fig:time_mqt_qchem}
\end{figure}

\emph{Clifford Insertion.}
\Cref{fig:time_mqt_qchem} reports the end-to-end duration effect of this optimization.
Clifford insertion primarily shortens the Clifford-only critical path. For end-to-end duration, it reduces circuit duration by 11.54\% on average on MQTBench and 2.97\% on average on QChem, with a maximum reduction of 32.41\% across the evaluated instances. The reduction is larger when measured on the Clifford-only portion of the critical path.
% For MQTBench, the duration of Cliffords can be reduced by up to 40\%, and the total duration can be reduced up-to 14\%.
% For QChem, the duration of Cliffords can be reduced by up to 19\%, while below 0.5\% for total duration.
% This result shows that Clifford insertion usefully shortens the Clifford portion of the circuit, but its end-to-end effect remains secondary compared with the failure-probability gains from syn@fac.

\emph{Logical qubit Allocation.}
We use the reproduced allocator of same-time work by Sethi \emph{et al.}~\cite{sethi_optimizing_2026} as the adopted backend throughout the main evaluation of the studied regime.
Note again, we do not position allocation as a contribution of this paper.
To isolate its effect, we perform an ablation that replaces the contiguous allocation used in Tour de Gross with the allocator, while keeping the remaining compilation choices identical to Tour de Gross.
For non-Clifford workloads, i.e., excluding Clifford-only cases where magic state teleportation is absent, this allocator alone reduces the estimated circuit failure probability by at most 5.4\% across the benchmark suite.
By contrast, \Cref{fig:breakdown_lpu_vs_fac} shows that the much larger reduction in failure probability comes from switching synthesis placement from syn@LPU to syn@fac.
Therefore, in the studied regime, inter-module measurement introduced by module communication is already given a strong backend treatment, yet it is still not the main driver of the headline failure-probability gains.

\section{Conclusion}
\label{sec:conclusion}

This paper is motivated by a general question for early modular FTQC:
what compiler optimizations most reduce the estimated circuit failure probability of realistic quantum algorithm workloads?
We answer that question for the studied BB system, namely the earliest nontrivial BB architecture of Tour de Gross with line connectivity, one factory, and up to three LPUs.
Within this studied system, and with the adopted allocator, synthesis placement yields the largest reduction in estimated circuit failure probability under the present cost model; syn@fac is favored because it reduces the teleportation of $\ket{T}$ states to LPUs.
Our later sensitivity analyses on cost model, LPU count, and factory count show that this conclusion is robust over a wider range of cost models and system sizes.
Based on our work, two future directions would keep on reducing the estimated circuit failure probability: reducing the logical error rate of inter-module measurement, and the cost of arbitrary-angle rotations.
We hope this work inspires the studies on compiler optimizations for early modular FTQC systems.
\begin{acknowledgments}
We thank Nikolaj Moll and Shifan Xu for valuable insights and discussions.
This work is supported by Boehringer Ingelheim, AFOSR MURI (FA9550-26-1-B036), and DARPA (under contract HR0011-26-9-E123). YD acknowledges partial support by the National Science Foundation (under awards OSI-2435244, CCF-2312754 and CCF-2338063), by the U.S. Department of Energy, Office of Science, National Quantum Information Science Research Center, Co-design Center for Quantum Advantage (C2QA) under Contract No. DE-SC0012704, by QuantumCT (under NSF Engines award ITE-2302908). External interest disclosure: YD is a consultant and equity holder for D-Wave Quantum, Inc.
\end{acknowledgments}
\appendix
\section{Additional Details of Background}
\label{app:background}

This appendix provides the full constructions behind the condensed background in \Cref{sec:pbc,sec:modular_compile,sec:bb_meas}.

\subsection{BB Codes}

BB codes can be described using cyclic shift matrices.
% Let $S_\ell$ be the $\ell \times \ell$ matrix that shifts basis states forward
Denote $S_\ell$ as the cyclic shift matrix of dimension $\ell \times \ell$,
where each row $i$ of the matrix has a $1$
at column $i+1$ modulo $\ell$.
Define $x = S_\ell$ and $y = S_m$.
Polynomials in $x$ and $y$ represent two-dimensional shifts,
e.g., $x^p y^q = S_\ell^p S_m^q$.
A BB code picks two polynomials $A = A_1+A_2+A_3$ and $B = B_1+B_2+B_3$ built
from monomials, and sets
$H^X = [A\ |\ B]$,
$H^Z = [B^\top\ |\ A^\top]$.
The instance used in this work is $[[144,12,12]]$ with parameters
$\ell = 12$, $m = 6$, $A = 1 + y + x^3 y^{-1}$, and
$B = 1 + x + x^{-1} y^{-3}$.
These dense logical memories come with more intricate logical operations.

\subsection{Detailed PBC-to-BB Lowering Gadgets}

This subsection gives the full gadget-level derivation behind the condensed lowering story in \Cref{sec:modular_compile}.
We now explain the circuit-level transformation from \Cref{fig:pipeline}(c) to \Cref{fig:pipeline}(d) using \Cref{fig:modular_compile}.

\textbf{Circuit language.}
\Cref{fig:modular_compile} uses four basic symbols: unitary gates, Clifford
rotations, Pauli measurements, and \emph{projections}, as shown in its legend.
A projection of $P$ means “measure $P$ and, if the outcome is $1$, apply a Pauli
correction $Q$ that anticommutes with $P$; if the outcome is $0$, do nothing.”
After projection of $P$,
we enforce the quantum state to be the $+1$ eigenstate of $P$.
Because Pauli corrections can be tracked in software, we don't materialize $Q$.

\textbf{Single-module primitives.}
Recall that Pauli-based computation needs only two ingredients:
arbitrary-angle Pauli rotations and Pauli measurements,
as shown in \Cref{fig:modular_compile}(a).
Both are implemented by a controlled-$P$ from an ancillary qubit onto target qubits.
We call the ancillary qubit the \emph{pivot} qubit, and 
the remaining qubits the \emph{compute qubits}.
The whole bundle of pivot and compute qubits is
an \emph{logical processing unit} (LPU).

\textbf{Multi-module operations.}
To implement a multi-module \( P(\varphi) \),
e.g., spanning from LPU-1 to LPU-3 in brown region of \Cref{fig:modular_compile}(b),
we still follow the template in \Cref{fig:modular_compile}(a).
Now, the pivot qubit on LPU-1 is the ancillary qubit for \( P(\varphi) \).
The corresponding parts of \( P(\varphi) \) on each LPU are \( P_1, P_2, P_3 \),
respectively.
The controlled-$P$ across modules
can be implemented by a GHZ preparation 
on pivot qubits of each module,
followed by local controlled-$P_i$s, (i = 1, 2, 3),
as shown in \Cref{fig:modular_compile}(b).
Multi-module Pauli measurements are implemented by the same template except that 
we perform a Pauli $X$ measurement on the pivot qubit of the first module.
Using the gadgets in \Cref{fig:modular_compile}(c),
we convert the intermediate circuit in \Cref{fig:modular_compile}(b) into the Pauli-based computation circuit in \Cref{fig:pipeline}(d).
Note that the $-P_i$ in \Cref{fig:modular_compile}(c) can be commuted
to the end and merged with the final measurement.

The resulting circuit, \Cref{fig:pipeline}(d), 
highlights which operations are not native to the chosen
code family by marking them with “*”.
The remaining subsections describe how we synthesize those non-native rotations
and measurements.

\subsection{Detailed In-Module Measurement Synthesis}

This subsection details the inherited in-module synthesis machinery summarized in \Cref{sec:bb_meas}.

For gross code,
extractors with practical cost have been developed~\cite{yoder_tour_2025},
and they expose a small generating set
of Pauli measurements $\mathcal{M} = \langle X_0, X_6, Z_0, Z_6 \rangle$.
BB codes support fault-tolerant shift automorphism unitaries $\mathcal{A}$, 
which consist of physical SWAP
gates with specific patterns to realize logical CNOTs.
We leave the details of shift automorphism unitaries to~\cite{yoder_tour_2025}.
We set qubit 0 as the pivot qubit.
Before and after a Pauli measurement $P$ enabled by the extractor,
we apply a shift automorphism unitary $U$ and its inverse $U^\dagger$, respectively.
The resulting \( U\, P\, U^\dagger \) is also a Pauli measurement.
This is a common technique to enlarge synthesizable Pauli measurement sets in PBC~\cite{peres_quantum_2023}.
Together they yield the \emph{native measurement} set
$
\mathcal{N} = \{U\, P\, U^\dagger \mid P \in \mathcal{M}, U \in \mathcal{A}\}.
$
From native measurements, we can synthesize Clifford rotations
by using the pivot qubit as an ancilla.
To implement \( R(\pi/4) \),
we require a native measurement, e.g., \( Q \otimes R \), 
that has non-identity support on the pivot qubit (\( Q \in \{X,Y,Z\} \)).
An example of native rotation when \( Q = X \) is shown in round box in \Cref{fig:meas_and_rot_synth}(b2),
using 3 native measurements.
The native rotation set is defined as
$
\mathcal{R} = \{R(\pi/4) \mid Q \otimes R \in \mathcal{N},\ Q \in \{X,Y,Z\}\}.
$

The \emph{native (Clifford) rotations} generate the Clifford group~\cite{yoder_tour_2025}.
Any Pauli string $P_i \in \mathcal{P}_{k-1}=\mathcal{P}_{11} := \{I,X,Y,Z\}^{\otimes 11}$ can be synthesized by sandwiching a
native measurement $Q \otimes R$ between native rotations:
$
P_i = (R_1 \dots R_k) \, R \, (R_k^\dagger \dots R_1^\dagger).
$
\Cref{fig:meas_and_rot_synth}(b1) implements this pattern.
When implementing native rotations, 
the pivot qubit serves as an ancilla and its state will be destroyed.
So all such rotations $R_i$ should be done before and after
the center native measurement (i.e., done in the gray region in \Cref{fig:meas_and_rot_synth}(b1)).

To estimate the synthesis cost we initialize $S$ with $\mathcal{N}$, then iteratively conjugate every element of
$S$ by every native rotation.
Whenever a new Pauli string appears we insert it into $S$ and continue.
The process stops once $S$ spans all $4^{11}$ Pauli strings besides $I^{\otimes 11}$, and
the count of \#native measurements
are shown in \Cref{tab:bb_in_mod_meas_count}.

\section{Additional Details of Clifford Insertion}
\label{sec:app_cliff_insertion}

This appendix gives the full formulation of the Clifford-insertion optimization summarized in \Cref{sec:depth_opt}.

Clifford insertion ideas exist in surface-code literature where Cliffords are statically inserted to remove $Y$-basis measurements because they are more expensive than $X$- and $Z$-basis measurements~\cite{stein_architectures_2024}.
However, this cannot be applied directly to BB codes since they have distinct Pauli measurement synthesis than surface codes.

Instead, we propose a \emph{dynamic} optimization that tailors to the Pauli measurement synthesis pattern where the cost could be described by a function $f: \mathcal{P}_{n} \to \mathbb{R}^+$ for $n$ qubits in a module.
This also covers the surface-code optimization, where the compiler scans the sequence from left to right and inserts Cliffords for each $Y$-basis measurement it sees~\cite{stein_architectures_2024}.

Note that neither we nor the contemporaneous work~\cite{sethi_optimizing_2026} introduced in-module qubit mapping in this paper.
First, brute-force permutation of all possible in-module qubit mappings for each module is not scalable.
Second, unlike qubit mapping in gate-based computation, there is no native SWAP-based strategy for BB codes.
However, Clifford is essentially a basis change of Pauli-based computation, which can intuitively be regarded as runtime logical ``SWAP'' in PBC.
From this perspective, this optimization is a kind of in-module dynamic qubit mapping.

\subsection{Clifford Insertion}

Recall that in \Cref{sec:modular_compile}, we introduced the modular PBC compilation flow used in this paper.
To optimize the duration of a PBC, we first build a directed acyclic graph (DAG) by representing each operation as a node and its temporal dependencies as directed edges.
This is shown in \Cref{fig:pipeline}(d).
Notice that we omit the measurements on pivot qubits for simplicity and efficiency, since they will not be affected by the optimization.
When computing the duration, we will add them back.
For each node, we assign the duration of the operation as the weight of the node.
Then we can find the longest-duration path in the DAG, which is the critical path, as shown in \Cref{fig:segment}.
On the critical path, we identify the \emph{segments}, defined as the longest consecutive sequence of operations on compute qubits in the same module.
A segment is allowed to span inter-module $ZZ$ measurements, since inter-module $ZZ$ measurements do not affect the optimization and their duration cannot be optimized either.
An example is shown in \Cref{fig:segment}, where the segment traverses the $ZZ$ measurement between \( YP_2' \) and \( YP_3' \).

Define a function $f: \mathcal{P}_{n} \to \mathbb{R}^+$ as the cost function of synthesizing a measurement $P$, and $n$ is the number of compute qubits in a module.
Define a segment as $S = \{P_1, P_2, \dots, P_l\}$.
We can convert the segment to another one with the same logical behavior by inserting Cliffords.
For example,
\begin{align*}
    P_1\,P_2\,P_3 &= (C^\dagger\,C)\,P_1\,(C^\dagger\,C)\,P_2\,(C^\dagger\,C)\,P_3\,(C^\dagger\,C) \\
    &= C^\dagger\,(C\,P_1\,C^\dagger)\,(C\,P_2\,C^\dagger)\,(C\,P_3\,C^\dagger)\, C.
\end{align*}
This means we can first implement \( C^\dagger \), then the new segment $\{C\,P_i\,C^\dagger\}_{i=1}^3$, and finally \( C \).
An intuition for this optimization is to find a shared axis rotation $C$ for all the rotations in the segment.
If the total cost of implementing the transformed segment \( \{C P_i C^\dagger\} \) plus the cost of implementing $C$ and $C^\dagger$ is less than the cost of the original sequence, the Clifford insertion is worthwhile.
Formally, for a segment, we want to solve
$
\max_{C} -f(C) - f(C^{\dagger}) + \sum_{i=1}^l f(P_i) - f(C\,P_i\,C^{\dagger}).
$
Theoretically, we can pick any cost function $f$.

\begin{table}[t]
    \centering
    \begin{tabular}{c|c|c|c|c|c}
        \hline
        \#Native meas. & 1 & 7 & 13 & 19 & 25 \\
        \hline
        Count & 245 & 12579 & 490770 & 3505249 & 185460 \\
        \hline
    \end{tabular}
    \caption{
        In-module native measurements required to synthesize arbitrary Pauli strings besides $I^{\otimes 11}$ in $\mathcal{P}_{11}$ for the BB codes studied here.
    }
    \label{tab:bb_in_mod_meas_count}
\end{table}

For BB codes, we define $f: \mathcal{P}_{11} \to \set{1, 7, 13, 19, 25}$ as the number of in-module measurements to synthesize a measurement $P$ according to \Cref{tab:bb_in_mod_meas_count}.

The same optimization template can also cover related surface-code settings.
Implementing $Y$-basis measurements on surface codes requires twist defects~\cite{chamberland_circuit-level_2022} that are not natively supported by some layouts.
A solution is to decompose $Y$-basis measurements into $X$/$Z$ measurements and $Z$/$X$ Clifford rotations~\cite{kan_sparo_2025}.

\subsection{Multi-Window Selection}

For a segment, if we use only one $C$ to conjugate all operations, the longer the segment is, the harder it becomes to reduce the overall cost of the segment.
Here we propose an algorithm to select multiple \emph{windows} to maximize the reduction for the segment.
An intuitive metaphor for this algorithm is buying and selling stocks with transaction fees.

The operation indices in the segment are $[1,L]$.
Define $\Delta_k[i] := f(P_i) - f(C_k\,P_i\,C_k^{\dagger})$ as the \emph{raw} cost reduction at position $i$ if $C_k$ is applied to $P_i$.
For BB codes, the native rotations $\mathcal{R}$ are cheap in terms of duration, so we insert them.
For a fixed library of Cliffords $\set{C_k} = \mathcal{R}$, choosing a window incurs a fixed overhead $s := f(C_k) + f(C^{\dagger}_k)$.
Denote a window by $(a,b,k)$, where $a$ is the starting position and $b$ is the ending position, and $C_k$ is the Clifford that opens and closes the window at $a$ and $b$, respectively.
Define $\text{dp}[i]$ as the best \emph{net} reduction using indices in $[1,i]$ such that all windows are disjoint and closed.
The goal is to find a set of disjoint windows that maximizes the net reduction
$$
\sum_{\text{windows}\,(a,b,k)} \big(\sum_{i=a}^b \Delta_k[i] - s\big)
$$
using a one-pass dynamic programming algorithm.
First, we precompute the prefix sum of $\Delta_k[i]$ as $\text{pref}_k[i] := \sum_{t=1}^i \Delta_k[t]$, i.e., accumulated raw reduction if $C_k$ is applied from $1$ to $i$.
The key observation is that, at position $i$, closing a window $w = (j,i,k)$ that uses $C_k$ and starts at position $j$ yields a net reduction of
\begin{equation}
    \label{eq:window_net_reduction}
\underbrace{\text{pref}_k[i] - \text{pref}_k[j-1]}_{\text{raw reduction of } w=(j,i,k)} - s +
\underbrace{\text{dp}[j-1]}_{\text{best net reduction before } w},
\end{equation}
which is rearranged to
\begin{equation}
    \label{eq:window_net_reduction_rearranged}
\text{pref}_k[i] + \underbrace{\text{dp}[j-1] - \text{pref}_k[j-1] - s}_{\text{``base''}},
\end{equation}
where $\text{pref}_k[i]$ is fixed while the terms in the bracket depend on $j$.
We call the term in the bracket the \emph{base}.
So, for each $k$, maximizing \Cref{eq:window_net_reduction} is equivalent to finding the best base by varying $j$:
$$
\text{base}_k[i] := \max_{1\leq j \leq i} \left( \text{dp}[j-1] - \text{pref}_k[j-1] - s \right).
$$
This asks: if we are determined to close a window with $C_k$ at position $i$, what is the best starting position $j$ that maximizes the base?
Then, we can compute the best net reduction from closing a window with $C_k$ at position $i$ in $O(1)$ as
$$
\text{close}_k[i] := \text{base}_k[i] + \text{pref}_k[i].
$$
The recurrence is (a) skip $i$: $\text{dp}[i] = \text{dp}[i-1]$, or (b) close some window that uses $C_k$ and maximizes the reduction.
So,
$$
\text{dp}[i] = \max(\text{dp}[i-1], \max_k \text{close}_k[i]).
$$
That is, among all possible windows that can be closed at position $i$, we choose the one that maximizes the net reduction.
Last, we update $\text{base}_k$ for the next iteration:
$$
\text{base}_k[i+1] = \max(\text{base}_k[i], \text{dp}[i] - \text{pref}_k[i] - s).
$$
The best net reduction is $\text{dp}[L]$.
The time complexity is $O(K L)$ and space complexity is $O(K L)$, where $K$ is the number of candidate Cliffords and $L$ is the length of the segment.
We can recover the best windows by backtracking.

We apply the multi-window selection algorithm to each segment of the critical path.
Next, we find the top $T$ segments with the largest net reduction in duration and apply Clifford insertion to these segments.
We then recompute the critical path because it may have changed.
We repeat this process until the duration reduction rate compared to the previous iteration is less than a threshold of $1\%$.

% \nocite{*}

\bibliography{ref_kun}% Produces the bibliography via BibTeX.

\end{document}